\documentclass[letterpaper, 10 pt, conference]{ieeeconf}  

\IEEEoverridecommandlockouts                              
\makeatletter
\let\NAT@parse\undefined
\makeatother

\usepackage[dvipsnames]{xcolor}

\newcommand*\linkcolours{ForestGreen}

\usepackage{times}
\usepackage{graphicx}
\usepackage{amssymb}
\usepackage{gensymb}
\usepackage{amsmath}
\usepackage{breakurl}

\usepackage{url,hyperref}
\hypersetup{
colorlinks,
linkcolor=\linkcolours,
citecolor=\linkcolours,
filecolor=\linkcolours,
urlcolor=\linkcolours}

\usepackage{algorithm}
\usepackage{algorithmic}

\usepackage[labelfont={bf},font=small]{caption}
\usepackage[none]{hyphenat}

\usepackage{mathtools, cuted}

\usepackage[noadjust, nobreak]{cite}

\usepackage{tabularx}
\usepackage{amsmath}

\usepackage{float}

\usepackage{pifont}

\newcommand*\GitHubLoc{https://github.com/Jeffrey-Ede/DLSS-STEM}
\newcommand*\coverage{1/16, 1/25, 1/36, 1/64, and 1/100 px coverage}

\newcolumntype{Y}{>{\centering\arraybackslash}X}

\usepackage[]{placeins}

\newcommand\extraspace{3pt}

\usepackage{placeins}

\usepackage{tikz}
\newcommand*\circled[1]{\tikz[baseline=(char.base)]{
            \node[shape=circle,draw,inner sep=0.8pt] (char) {#1};}}
            
\usepackage[framemethod=tikz]{mdframed}

\usepackage{afterpage}

\usepackage{stfloats}

\usepackage{atbegshi}
\newcommand{\handlethispage}{}
\newcommand{\discardpagesfromhere}{\let\handlethispage\AtBeginShipoutDiscard}
\newcommand{\keeppagesfromhere}{\let\handlethispage\relax}
\AtBeginShipout{\handlethispage}

\usepackage{comment}

\usepackage{caption}
\DeclareCaptionLabelFormat{nospace}{#1 #2}
\renewcommand{\thetable}{\arabic{table}}

\title{\Large \bf
Deep Learning Supersampled Scanning Transmission Electron Microscopy
}

\author{Jeffrey M. Ede\vspace{0.5\baselineskip}\\
j.m.ede@warwick.ac.uk
}

\begin{document}

\maketitle
\thispagestyle{empty}
\pagestyle{empty}

\begin{abstract}

Compressed sensing can increase resolution, and decrease electron dose and scan time of electron microscope point-scan systems with minimal information loss. Building on a history of successful deep learning applications in compressed sensing, we have developed a two-stage multiscale generative adversarial network to supersample scanning transmission electron micrographs with point-scan coverage reduced to 1/16, 1/25, ..., 1/100 px. We propose a novel non-adversarial learning policy to train a unified generator for multiple coverages and introduce an auxiliary network to homogenize prioritization of training data with varied signal-to-noise ratios. This achieves root mean square errors of 3.23\% and 4.54\% at 1/16 px and 1/100 px coverage, respectively; within 1\% of errors for networks trained for each coverage individually. Detailed error distributions are presented for unified and individual coverage generators, including errors per output pixel. In addition, we present a baseline one-stage network for a single coverage and investigate numerical precision for web serving. Source code, training data, and pretrained models are publicly available at \url{\GitHubLoc}.

\end{abstract}

\section{INTRODUCTION}

\noindent Materials can be damaged by an electron beam\cite{wolf2018stem, garcia2014analysis} as they are examined by scanning electron microscopy\cite{sem_review} (SEM), scanning transmission electron microscopy\cite{tong2018scanning} (STEM), and other point-scan systems. This limits materials that can be studied at high electron doses to stable crystals and select organic structures, and is particularly problematic at high resolution as local electron dose may be higher. In this context, we have developed deep learning supersampling for STEM (DLSS-STEM) with a generative adversarial network\cite{goodfellow2014generative, wang2017high} (GAN) to increase resolution, decrease scan time and lower electron dose.

There has been increased interest in compressed sensing in electron microscopy\cite{binev2012compressed}, especially over the last 5-10 years. Recent advances include the completion of STEM images from spiral scans\cite{jmede2019partialscan, li2018compressed}, TEM video inpainting\cite{reed2019electrostatic}, and interpolative SEM\cite{anderson2013sparse} and STEM\cite{sanders2018inpainting} inpainting. In particular, compressed sensing is enabling new biological applications\cite{ferroni2016biological, guay2016compressed} by decreasing electron beam damage.

Deep learning has a history of successful applications to image infilling, including image completion\cite{wu2018deep}, irregular gap infilling\cite{liu2018image} and supersampling\cite{yang2018deep}. This has motivated the application of deep learning to electron microscopy supersampling. Recent advances include a network trained on pairs of low and high resolution STEM images\cite{de2019resolution}, and a network trained to generalize from artificial noise to increase the resolution limit of a SEM\cite{fang2019deep}. However, collecting a dataset of image pairs is time-intensive and may be complicated by alignment, and artificial noise decorrelates outputs from inputs.

\begin{figure}[tbh!]
\centering
\includegraphics[width=0.85\columnwidth]{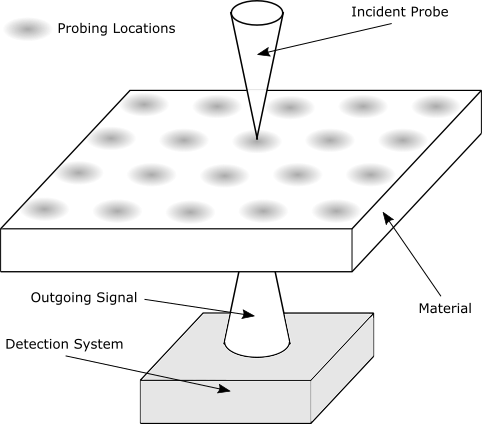}
\caption{ Simplified point-scan system showing separation between probing locations. }
\label{stem_spots}
\end{figure}

We have been developing a simpler approach, which uses deep learning to supersample STEM images after nearest neighbour downsampling. This creates examples from single images, and outputs are correlated with inputs. A simplified STEM raster scan system is illustrated in fig.~\ref{stem_spots} to show that STEM pixel intensities are measured at a regular lattice of probing locations. It follows that if a STEM image is nearest neighbour downsampled to have probing location coverage $c$, $c \in \{1/s^2 \,|\, s \in \mathbb{N} \}$, of its original coverage, it is physically equivalent to a lower resolution scan. This provides an inexpensive method to generate low and high resolution image pairs at multiple coverages.



\section{Experiment}

\noindent To train a neural network for arbitrary STEM images, we used 161069 512$\times$512 32 bit floating point STEM crops from the STEM Crops dataset\cite{warwickem!} introduced in \cite{jmede2019partialscan}. The dataset is collated from individual micrographs saved to University of Warwick data servers by dozens of scientists working on hundreds of projects and therefore has a diverse constitution. We use the default split into 110933 training, 21259 validation, and 28877 test crops, where each set is shuffled independently. Each set is collected by a different subset of scientists and has different characteristics. As a result, unseen validation and test sets can be used to quantify the ability of a trained network to generalize.

Each 512$\times$512 crop was linearly transformed to have a minimum value of 0 and maximum of 1, except for a small number of uniform images that were scaled to 0.5. Any non-finite values; such as $\pm\infty$ and NaN, were replaced with 0 before transformation. Each crop was then subject to a random combination of flips and 90$\degree$ rotations to augment the dataset by a factor of eight. Finally, each 512$\times$512 crop was nearest neighbour downsampled to either 1/16, 1/25, 1/36, 1/64, or 1/100 px coverage, depending on the experiment.

\begin{figure}[tb]
\centering
\includegraphics[width=0.97\columnwidth]{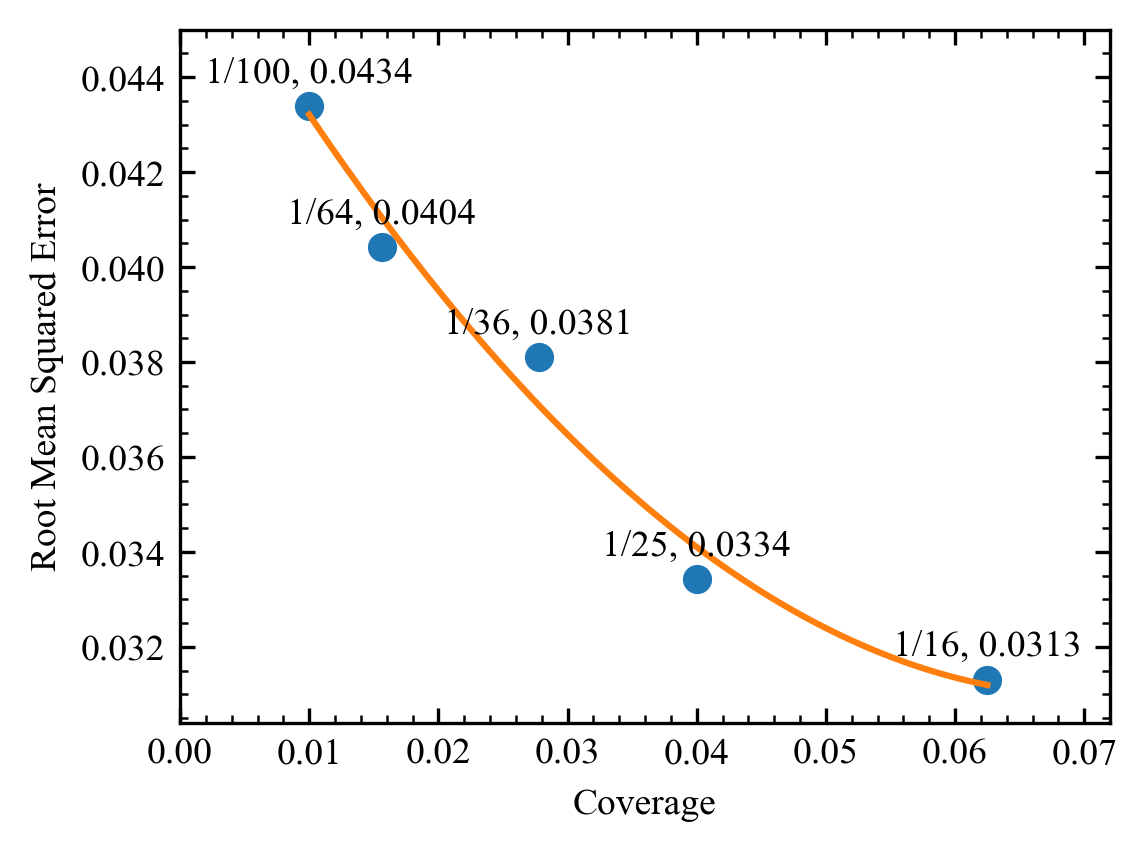}
\caption{ Root MSEs for 20000 training set crops in $[0, 1]$ after non-adversarial training for 10$^6$ iterations are fit by a quadratic. }
\label{training_errors}
\end{figure}

\begin{figure}[tbh!]
\centering
\includegraphics[width=0.8\columnwidth]{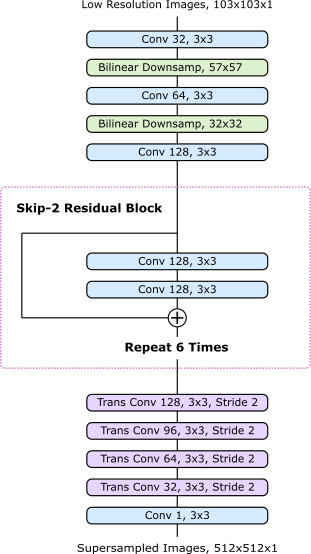}
\caption{ One-stage  generator  that  supersamples  low  resolution 103$\times$103$\times$1 images to 512$\times$512$\times$1. }
\label{gen-1-step}
\begin{tabular}{c}
\vspace{\baselineskip}\\
\end{tabular}
\includegraphics[width=0.97\columnwidth]{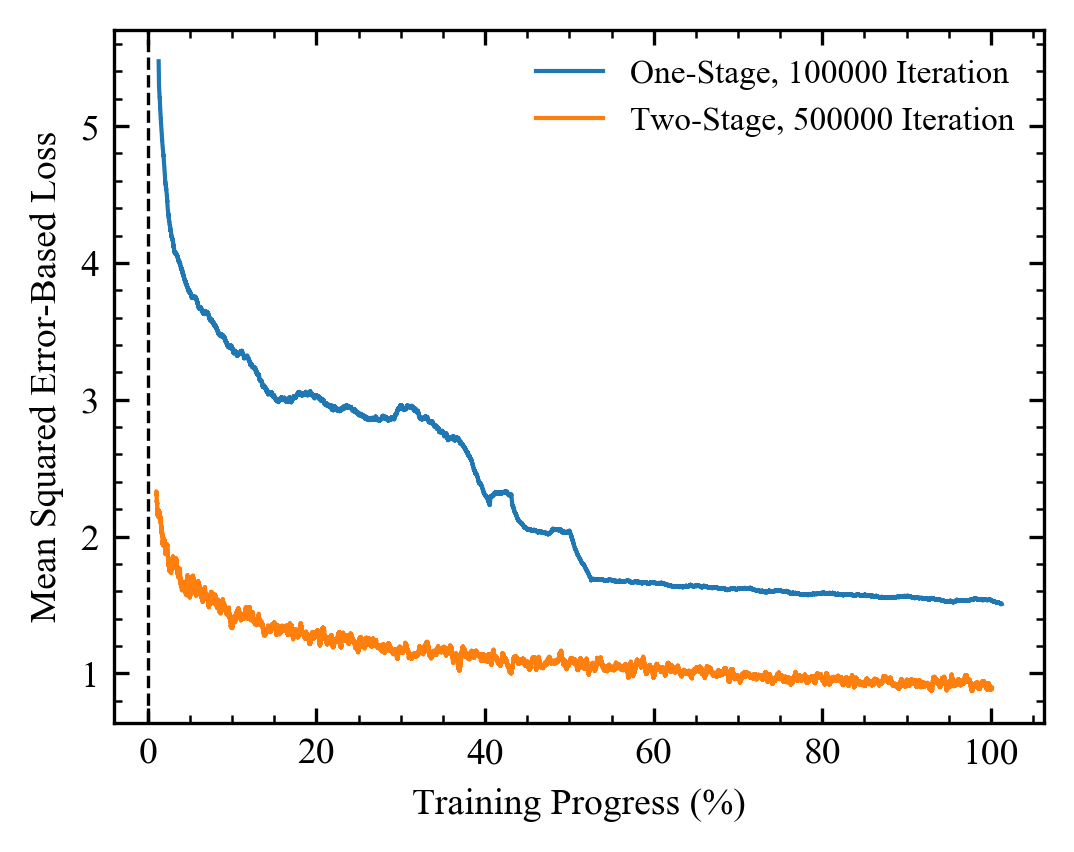}
\caption{ ALRC MSE losses for a 1/25 px coverage one-stage and two-stage generators, where batch normalization is frozen halfway through one-stage generator training. For clarity, the first 1\% of iterations before the dashed line, where losses rapidly decrease, are not shown and learning curves are 2500 iteration boxcar averaged. }
\label{learning_curves}
\end{figure}

Most of our experiments are based on a two-stage generator, ancillary trainer and multi-scale discriminator developed for 512$\times$512 partial-STEM\cite{jmede2019partialscan}; see figs.~\ref{gen-2-step}-\ref{discr} in section~\ref{sec:architecture}. Input crops were nearest neighbour upsampled to 512$\times$512, regardless of their size so a network could be trained for a range of pixel coverages. For a single coverage, nearest neighbour upsampling inefficiently utilizes initial generator convolutions. As a result, we also developed a one-stage generator shown in fig.~\ref{gen-1-step} and will show that it can achieve similar performance for a single coverage.

To train networks for a single coverage, we used the two-stage generator and non-adversarial learning policy in \cite{jmede2019partialscan}, other than the replacement of Huberisation with adaptive learning rate clipping\cite{ede2019adaptive} (ALRC). Final root mean square training errors for 1/16, 1/25, 1/36, 1/64, and 1/100 coverage are displayed in fig.~\ref{training_errors} and mean square square errors are fitted by the polynomial
\begin{equation}
p_\text{raw}(c) = 0.002211 - 0.037887c + 0.289451c^2.
\end{equation}
Mean squared errors increase as coverage decreases. As a result, a multiple coverage network will emphasise optimization for low coverage where losses are higher unless losses are adjusted. To scale MSEs so that all losses are on the same scale for coverages $c_\text{all} = \{ 1/s^2 \,|\, s \in [4, 10], \mathbb{N} \}$, we divide MSEs by
\begin{equation}
p(c) = \dfrac{ p_\text{raw}(c) }{ \text{mean}(\{p_\text{raw}(x) \,|\, x \in c_\text{all} \}) }.
\end{equation}
If $p(c)$ is not known in advance, $p_\text{raw}(c)$ could be tracked throughout training by an exponential moving average.

In detail, our generator, $G$, is partitioned into inner, $G_\text{inner}$, and outer, $G_\text{outer}$, generators, and an inner generator trainer, $T$, in fig.~\ref{gen-2-step}. The generator, $G(I) = G_\text{outer}(G_\text{inner}(I), I)$, for an input, $I$, and is all that is needed for inference. To reduce gradient variance introduced by varying noise levels during training by stochastic gradient decent, target crops were blurred to $I_\text{blur}$ by a 5$\times$5 symmetric Gaussian kernel with a 2.5 px standard deviation. The ALRC scaled MSE loss is
\begin{equation}\label{loss_G_MSE}
L_{G, \text{MSE}} = \text{ALRC}\left(\lambda_{G, \text{MSE}}  \frac{\text{MSE}(G(I),  I_\text{blur})}{p(c)}\right),
\end{equation}
where MSE($x_1$, $x_2$) is the MSE between $x_1$ and $x_2$, ALRC($x$) apples ALRC to $x$, and we chose $\lambda_{G, \text{MSE}} = 200$.

Gaussian blurring is a linear transformation that preserves image information, and can be reversed by deconvolution for inference. However, Fourier deconvolution is numerically unstable for small kernels as division by their Fourier transforms is denominated by small numbers. Instead, we find that deconvolution by ADAM optimized gradient descent for 100 iterations, $i$, with learning rate $\eta = \eta_0 a^i$, where $\eta_0 = 0.3$ and $a=0.99$, and first moment of the momentum decay, $\beta=0.9$, could reduce the MSE to $6.28\times10^{-8}$ for 1000 random training set crops in $[0, 1]$. In comparison, the MSE between blurred and unblurred crops was $3.03\times10^{-3}$. Other, potentially faster, algorithms; such as ADMM\cite{thakallapelli2019alternating} and Wiener filtering\cite{chen2006new}, have also been developed for finite kernel deconvolution.

\begin{figure}[tb]
\centering
\includegraphics[width=0.45\columnwidth]{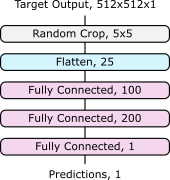}
\caption{ Multilayer perceptrons predict outer generator MSEs by examining noise characteristics of 5$\times$5 random crops from 512$\times$512 target outputs. }
\label{fig:dlss_mlp}
\end{figure}

To avoid deconvolution artefacts, it may be desirable to train with unblurred images. However, this prioritizes high noise examples as their MSEs will be higher. To give all examples the same priority, we propose training $N=20$ identical multilayer perceptrons, denoted $M$ and shown in fig.~\ref{fig:dlss_mlp}, to predict the outer generator MSE, from random $d \times d$, where $d=5$, crops, $\{C_1, C_2, ..., C_N\}$, to minimize
\begin{equation}
L_M = \frac{1}{N} \sum\limits_{i=1}^{N} \text{MSE}\left(M(C_i), \frac{\lambda_{G, \text{MSE}}\text{MSE}(G(I), I)}{p(c)}\right)
\end{equation}
where crops are chosen randomly, and $d \ll 512$ so that predictions are based on noise characteristics; rather than large-scale features. The mean prediction is
\begin{equation}
\mu_M = \frac{1}{N} \sum\limits_{i=1}^{N} M(C_i)
\end{equation}
and can used to homogenize prioritization by modifying the outer generator loss proposed in eqn.~\ref{loss_G_MSE} to 
\begin{equation}\label{loss_G_MSE_no_blur}
L_{G, \text{MSE}}^\text{no blur} = \text{ALRC}\left(\lambda_{G, \text{MSE}}  \frac{\mu_1\text{MSE}(G(I),  I)}{p(c)\max(\mu_M, \epsilon)}\right),
\end{equation}
where $\mu_1$ is an exponential moving average that tracks the first raw moment of the error, i.e. implemented for ALRC, and $\epsilon=0.1$ is a safeguard to prevent small $m$ producing extreme $L_{G, \text{MSE}}$ in the early stages of training. To increase stability, an exponential moving average of $M$ weights could be used to calculate $m$.

\begin{figure*}[htbp]
\footnotesize
Learning rate, $\eta$\\
\begin{tabular*}{\textwidth}{c@{\extracolsep{\fill}}cccc}
\hline
\multicolumn{1}{c}{Iterations, $i$} & Generator, $G$ & Predictor, $M$ & Trainer, $T$ & Discriminators, $D$ \\
\hline
$[0, 5\times10^5)$ & 0.0003 & 0.0003 & 0.0006 & - \\
$[5\times10^5, 1\times 10^6)$ & $0.0003\text{f}(i, 5\times 10^5)$ & $0.0003\text{f}(i, 5\times 10^5)$ & $0.0006\text{f}(i, 5\times 10^5)$ & - \\
$[1\times10^6, 1.5\times 10^6)$ & 0.0001 & - & 0.0002 & 0.0001 \\
$[1.5\times10^6, 2\times 10^6)$ & $0.0001\text{f}(i, 5\times 10^6)$ & - & $0.0002\text{f}(i, 5\times 10^6)$ & 0.0001 \\
\hline
\end{tabular*}
\vspace{\baselineskip}\\
Momentum decay, $\beta$\\
\begin{tabular*}{\textwidth}{c@{\extracolsep{\fill}}cccc}
\hline
\multicolumn{1}{c}{Iterations, $i$} & Generator, $G$ & Predictor, $M$ & Trainer, $T$ & Discriminators, $D$ \\
\hline
$[0, 5\times10^5)$ & 0.9 & 0.9 & 0.9 & - \\
$[5\times10^5, 1\times 10^6)$ & $0.5 + 0.4\text{f}(i, 5\times 10^5)$ & 0.9 & $0.5 + 0.4\text{f}(i, 5\times 10^5)$ & - \\
$[1\times10^6, 1.5\times 10^6)$ & 0.5 & - & 0.5 & 0.5 \\
$[1.5\times10^6, 2\times 10^6)$ & 0.5 & - & 0.5 & 0.5 \\
\hline
\end{tabular*}
\captionof{table}{ ADAM optimizer learning rates, $\eta$, and first moment of the momentum decay rate, $\beta$, at each training iteration, $i$, for a two-stage generator, inner generator trainer, multiscale discriminators, optional loss predictors. The function $\text{f}: i, b \rightarrow 1 - \text{floor}(8i/b-1)$ describes a linear decay schedule with seven steps. }
\label{ADAM_hyperparam}
\end{figure*}

Generated pixels at input pixel locations have the same values as input pixels. As a result, the generator will focus on outputting the known input pixels. To prevent this, we replace inputted output pixels with their target values so they do not contribute to the loss function. 

To provide a shorter path for gradients to backpropagate to the inner generator and an additional regularization mechanism, the inner generator trainer learns to output half-size, bilinearly downsampled blurred target outputs, $I_\text{blur}^\text{half}$. The inner generator trainer cooperates with the inner generator to minimize
\begin{equation}
L_{G, \text{aux}} = \text{ALRC}\left(\lambda_{G, \text{aux}} \frac{\text{MSE}(T(G_\text{in}(I)), I_\text{blur}^\text{half})}{p(c)}\right),
\end{equation}
where we chose $\lambda_{G, \text{aux}} = 200$.

The first half of training is based on non-adversarial MSEs. Optionally, the generator can be adversarially fine-tuned as part of a GAN to produce images with realistic noise characteristics. We use a multiscale discriminator, $D = \{D_1, D_2, D_3 \}$, where $D_1$, $D_2$, and $D_3$ are shown in fig.~\ref{discr} and process single random 70$\times$70, 140$\times$140 and 280$\times$280 crops. The adversarial discriminator loss is
\begin{equation}
L_D = \frac{1}{N} \sum_{i=1}^{N} D_i(G(I))^2 + (D_i(I_N) - 1)^2,
\end{equation}
where $N=3$ is the number of discriminator scales. The adversarial generator loss is 
\begin{equation}
L_{G, \text{adv}} = \frac{1}{N} \sum_{i=1}^{N} D_i(G(I) - 1)^2. 
\end{equation}
Added together, the total generator loss during adversarial fine-tuning is
\begin{equation}
L_G = L_{G, \text{MSE}}^\text{blur} + L_{G, \text{aux}} + L_{G, \text{adv}},
\end{equation}
where the superscript of $L_{G, \text{MSE}}^\text{blur}$ indicates that generator outputs, $G(I)$, in eqn.~\ref{loss_G_MSE} or eqn.~\ref{loss_G_MSE_no_blur} are blurred by a 3$\times$3 symmetric Gaussian kernel with 1.5 px standard deviation to reduce suppression of realistic noise characteristics during adversarial training.


Other than loss function modifications, learning policy is the same as \cite{jmede2019partialscan}. As a result, it is only stated in the remainder of this section for ease of reference. Detailed justification and experiments can be found in \cite{jmede2019partialscan}.

\vspace{\extraspace}
\noindent \textbf{Optimization}: Training is ADAM\cite{kingma2014adam} optimized for 10$^6$ non-adversarial iterations, optionally followed by 10$^6$ adversarial iterations. ADAM learning rate, $\eta$, and first moment of the momentum decay rate, $\beta$, are tabulated for each iteration for the two-stage generator, inner generator trainer, discriminators, and optional loss predictors in table~\ref{ADAM_hyperparam}. All two-stage generator training was performed with batch size 1.

\vspace{\extraspace}
\noindent \textbf{ALRC}: All ALRC\cite{ede2019adaptive} layers were initialized with first raw moment $\mu_1 = 25$, second raw moment $\mu_2 = 30$, exponential decay rates $\beta_1 = \beta_2 = 0.999$, and $n = 3$ standard deviations.

\vspace{\extraspace}
\noindent\textbf{Input normalization:} Generator inputs were linearly transformed to $[0, 1]$. Target generator outputs in $[0, 1]$ were linearly transformed to $[-1, 1]$ before being randomly cropped and input to discriminators.

\begin{figure*}[htbp]
\centering
\footnotesize
\includegraphics[width=\textwidth]{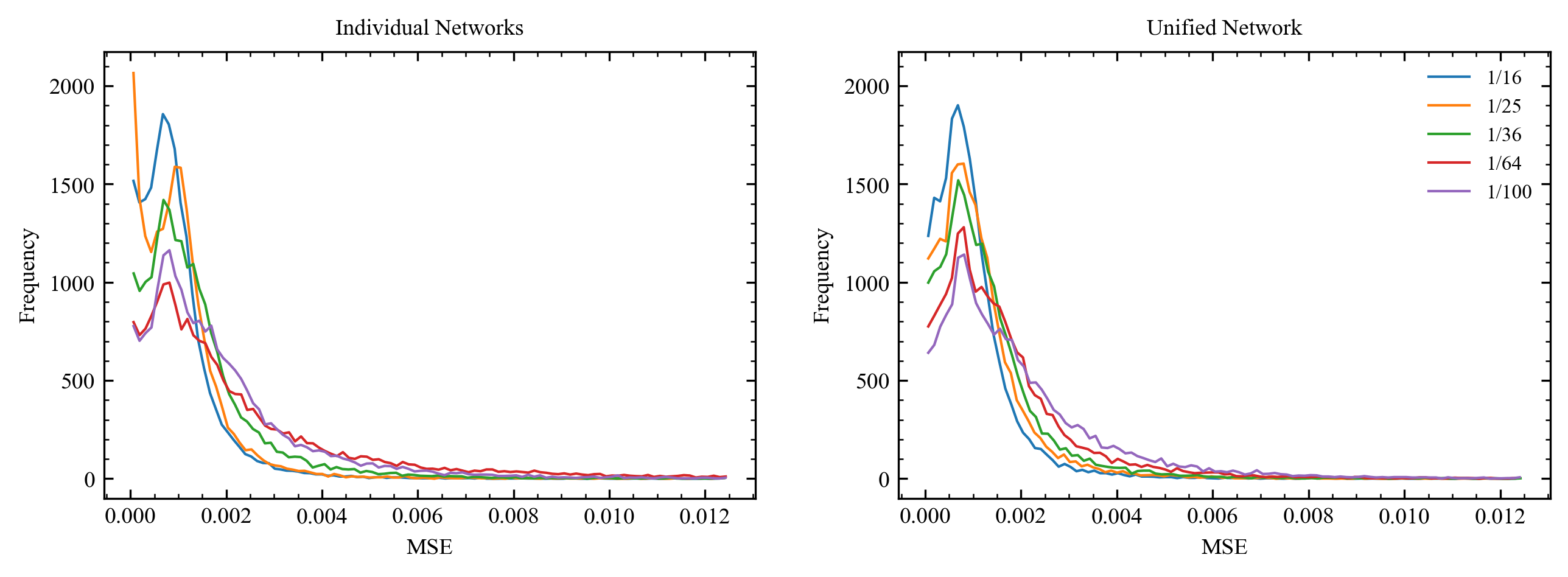}
\caption{ Frequency distributions of test set MSEs for 20000 images in $[0, 1]$ of individual and unified networks for \coverage. In coverage order, 60, 65, 88, 631, and 165 tail MSEs above 0.0125 are not shown for individual networks and 66, 62, 72, 115, and 142 are not show for the unified network. The remaining MSES in [0, 10) are distributed across 100 equispaced bins. }
\label{1_stage_vs_2_stage}
\begin{tabular}{c}
\vspace{\baselineskip}\\
\end{tabular}
\includegraphics[width=0.99\textwidth]{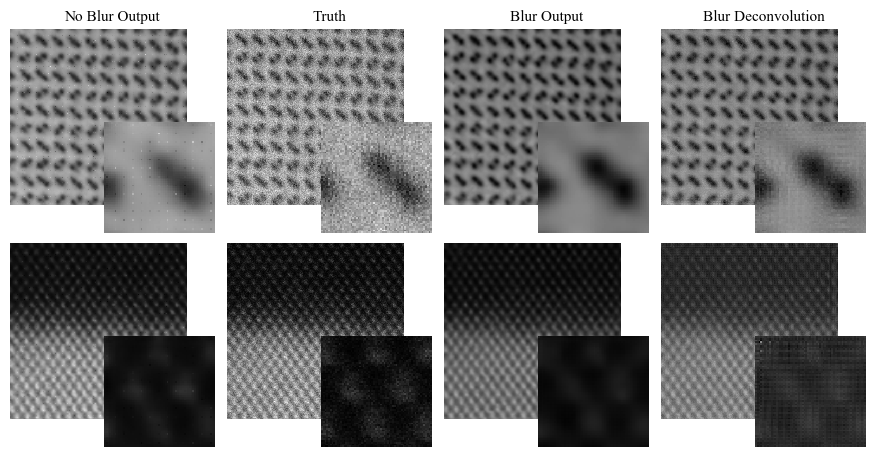}
\captionof{figure}{ Non-adversarial unified DLSS-STEM outputs for 1/49 coverage for target images with and without blurring by a 5$\times$5 symmetric Gaussian kernel with a 2.5 px standard deviation. Training with blurring produces blurrier outputs than training without blurring. Enlarged 64$\times$64 regions from the top left of each image reveal high-frequency grid-like spatial variation for Gaussian kernel deconvolution. True values replace no blur output values at input pixel locations. }
\label{deconv_panel}
\begin{tabular}{c}
\vspace{\baselineskip}\\
\end{tabular}
\begin{tabular*}{\textwidth}{l@{\extracolsep{\fill}}cccccccccc}
\hline
\multicolumn{1}{c}{} & \multicolumn{2}{c}{1/16} & \multicolumn{2}{c}{1/25} & \multicolumn{2}{c}{1/36} & \multicolumn{2}{c}{1/64} & \multicolumn{2}{c}{1/100} \\
Method & Mean & Std Dev & Mean & Std Dev & Mean & Std Dev & Mean & Std Dev & Mean & Std Dev \\
\hline
Nearest Neighbour & 0.0810 & 0.0641 & 0.0887 & 0.0720 & 0.0909 & 0.0745 & 0.0913 & 0.0749 & 0.0973 & 0.0815 \\
Area & 0.0810 & 0.0641 & 0.0811 & 0.0650 & 0.0844 & 0.0685 & 0.0913 & 0.0749 & 0.0932 & 0.0779 \\
Bilinear & 0.0542 & 0.0427 & 0.0609 & 0.0495 & 0.0634 & 0.0525 & 0.0654 & 0.0555 & 0.0703 & 0.0612 \\
Bicubic & 0.0687 & 0.0541 & 0.0745 & 0.0596 & 0.0769 & 0.0623 & 0.0788 & 0.0648 & 0.0836 & 0.0701 \\
Lanczos & 0.0728 & 0.0575 & 0.0786 & 0.0629 & 0.0810 & 0.0656 & 0.0829 & 0.0681 & 0.0877 & 0.0732 \\
\hline
DLSS-Individual & 0.0320 & 0.0427 & 0.0343 & 0.0447 & 0.0388 & 0.0492 & 0.0452 & 0.0800 & 0.0450 & 0.0535 \\
DLSS-Unified & 0.0323 & 0.0415 & 0.0346 & 0.0442 & 0.0372 & 0.0462 & 0.0417 & 0.0503 & 0.0454 & 0.0521 \\
\hline
\end{tabular*}
\captionof{table}{ Root mean and root standard deviations of supersampling mean squared errors for 20000 test set images in $[0, 1]$ after nearest neighbour downsampling to \coverage. DLSS-Individual results are for networks trained for each coverage individually whereas DLSS-Unified results are for a network trained for coverages $1/s^2, s \in [4, 10], \mathbb{N}$. }
\label{supersampling_methods}
\vspace{0.5cm}
\end{figure*}

\vspace{\extraspace}
\noindent\textbf{Weight normalization:} All generator weights are weight normalized and running mean-only batch normalization \cite{salimans2016weight, hoffer2018norm} is applied to the output channels of every convolutional layer, except the last. Channel means are tracked by exponential moving averages with decay rates of 0.99 and frozen in the second half of training.

\vspace{\extraspace}
\noindent \textbf{Spectral normalization:} All discriminator weights are spectrally normalized\cite{miyato2018spectral}. We use the power iteration method with one iteration per training step to enforce a spectral norm of 1 for each weight matrix. 

\vspace{\extraspace}
\noindent\textbf{Activation:} In the generator, ReLU\cite{nair2010rectified} non-linearities are applied after running mean-only batch normalization. In the discriminators, slope 0.2 leaky ReLU\cite{maas2013rectifier} non-linearities are applied after every convolution layer.

\vspace{\extraspace}
\noindent\textbf{Initialization:} Generator weights were initialized from a normal distribution with mean 0.00 and standard deviation 0.05. To apply weight normalization, a 512$\times$512 image with values sampled from a uniform distribution over $[-0.8, 0.8]$ is then propagated through the network. Each layer output is divided by its L2 norm and the layer weights assigned their division by the square root of the L2 normalized output's standard deviation. There are no biases in the generator.

Discriminator weights were initialized from a normal distribution with mean 0.00 and standard deviation 0.03. Biases were zero initialized.

\vspace{\extraspace}
\noindent \textbf{Experience replay:} An experience replay kept examples with errors in the top 20\%. In each iteration, there was a 20\% probability than an experience would be replayed.

\begin{figure}[b]
\centering
\includegraphics[width=\columnwidth]{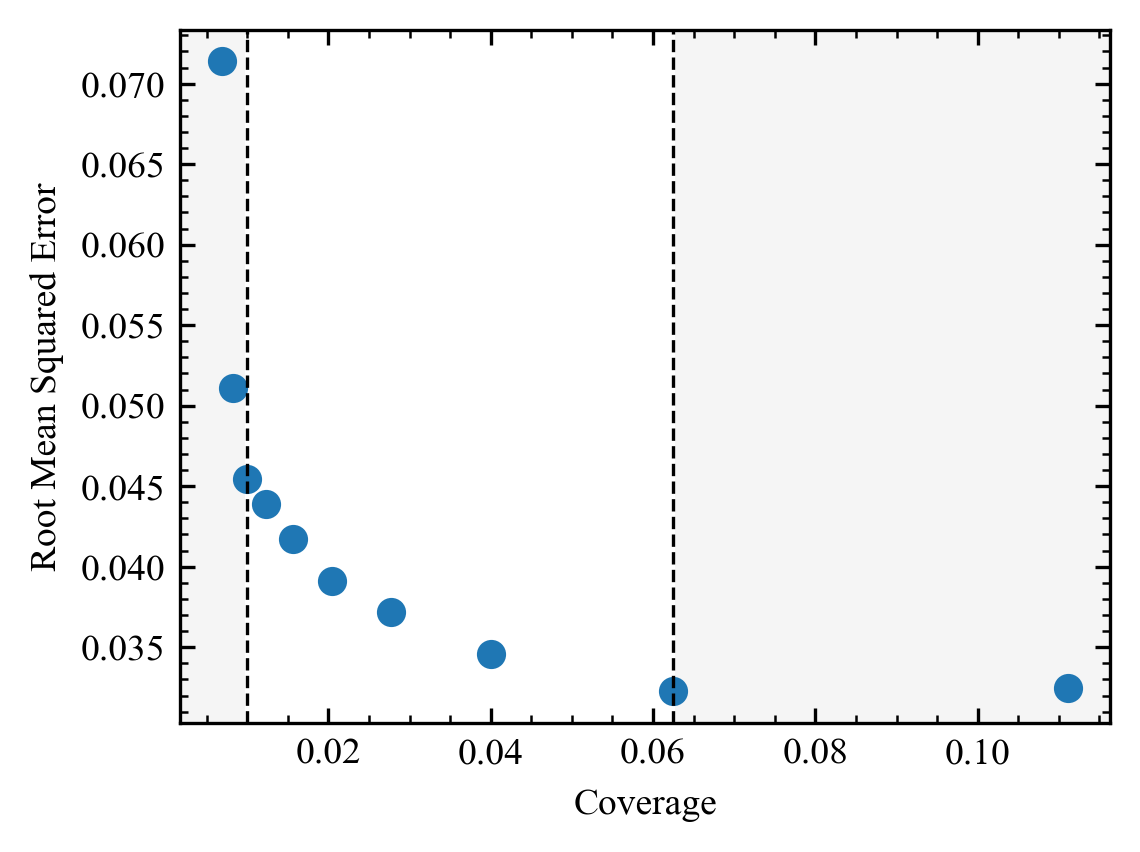}
\caption{ Root MSEs for 20000 test set crops in $[0, 1]$ after non-adversarial training a unified network for $10^6$ iterations. The unified network is trained for coverages in $[1/100, 1/16]$ px, between the dashed lines, and has high errors for unseen coverages. }
\label{test_errs}
\end{figure}

\begin{figure}[tbh!]
\centering
\includegraphics[width=0.8\columnwidth]{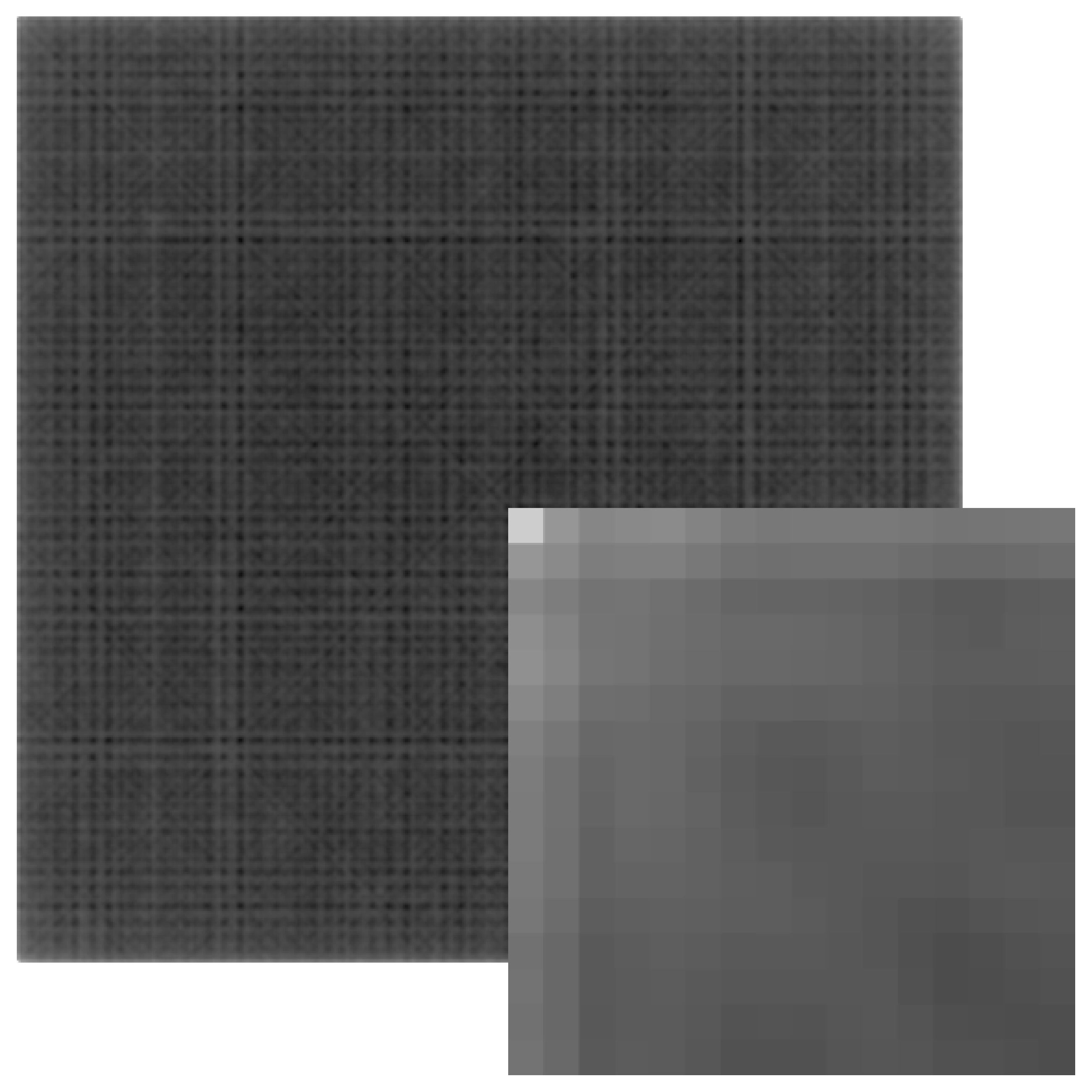}
\caption{ DLSS-Unified root MSEs at each 512$\times$512 generator output pixel for 20000 test set images. An enlarged 16$\times$16 region from the top-left is inset to show that root MSEs are higher near edges. }
\label{output_mse_image}
\vspace{0.5\baselineskip}
\includegraphics[width=\columnwidth]{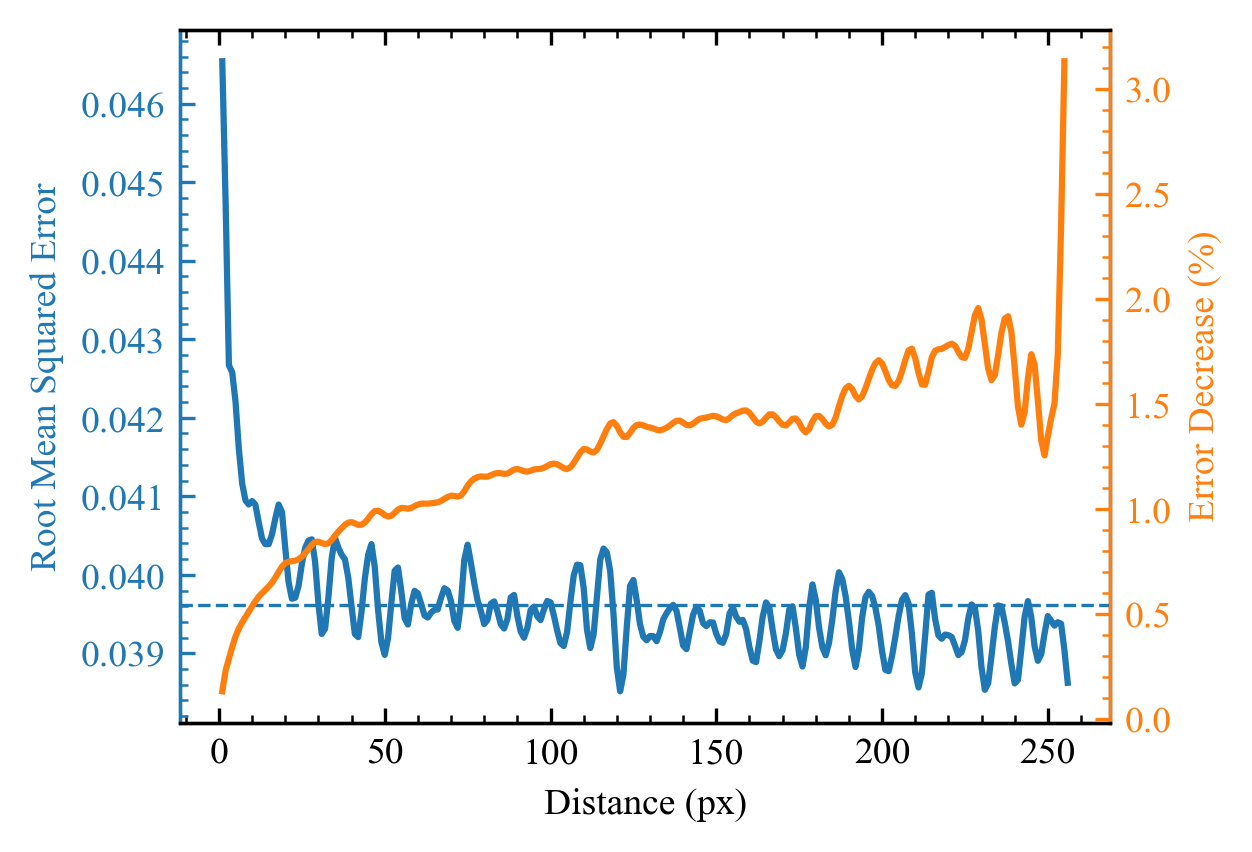}
\caption{ Root MSEs decrease with increasing minimum distance from generator output edges. Error oscillations are a result of systemic spatial error variation and a dashed line indicates the mean error. Mean errors decrease if pixels below distance thresholds from the edges are discarded; however, fewer pixels inform values at higher distances. }
\label{output_mse_dist}
\end{figure}

\section{Performance}

\noindent Individual and unified network MSEs for 20000 test set crops at \coverage are compared against MSEs for traditional methods in table~\ref{supersampling_methods}. In-line with training, target images were Gaussian blurred by a 5$\times$5 Gaussian kernel with a 2.5 px standard deviation. Individual and unified networks outperform all other methods; however, some algorithms are not designed for noisy images so some comparisons may be uncharitable.

DLSS-Unified root MSEs are $\sim$1\% higher than DLSS-Individual root MSEs, except for an outlier 1/36 px coverage DLSS-Individual root MSE where the DLSS-Unified root MSE is $\sim$4\% lower. In contrast, DLSS-Unified root standard deviations of MSEs are $\sim$2\% lower. Higher unified MSEs demonstrate that learning multiple; rather than individual, coverages decreases performance. However, lower unified standard deviations also suggest that multiple coverages regularize the network; making it more robust to high MSEs.

Frequency distributions of 20000 test set MSEs for unified and individual networks are shown in fig.~\ref{1_stage_vs_2_stage}. Most networks have a frequency peak at a MSE of $\sim$0.001, which decreases as coverage decreases and the portion of higher MSEs increases. These peaks may be the result of potions of images without significant fine detail below average pixel separations in low coverage crops. Individual networks also have peaks near a MSE of 0 for blank or otherwise uniform images that is not present for unified networks. The absence of this peak partly accounts for slightly higher DLSS-Unified MSEs and suggests that individual networks may be better at recognising some input classes.

Unified network test set root MSEs for training coverages, $1/s^2, s \in [4, 10], \mathbb{N}$, and unseen 1/9, 1/121, and 1/144 px coverages are displayed in fig.~\ref{test_errs}. The unified network is able to generalize; however, MSEs for unseen coverages are higher than might be extrapolated from seen coverages. In part, high errors for unseen coverages may be the result of nearest neighbour upsampling inputs; allowing generators to tailor initial convolutional kernels for various size uniform square regions. If so, bilinear or other smooth upsampling may improve the ability of generators to generalize.

Root MSEs of DLSS-Unified generator output pixels for 20000 test set crops are displayed in fig.~\ref{output_mse_image} for coverages $1/s^2, s \in [4, 10], \mathbb{N}$. Varied intensity lines show systematic spatial error variation, and errors are especially high at output edges. Root MSEs decrease with increasing minimum distance to image edges in fig.~\ref{output_mse_dist}. Root MSEs can be reduced by $\sim$1\% by discarding pixels close to the edges; however, returns diminish after discarding pixels less than $\sim$20 px from the edges.

\begin{figure*}[tbp!]
\centering
\includegraphics[width=0.97\textwidth]{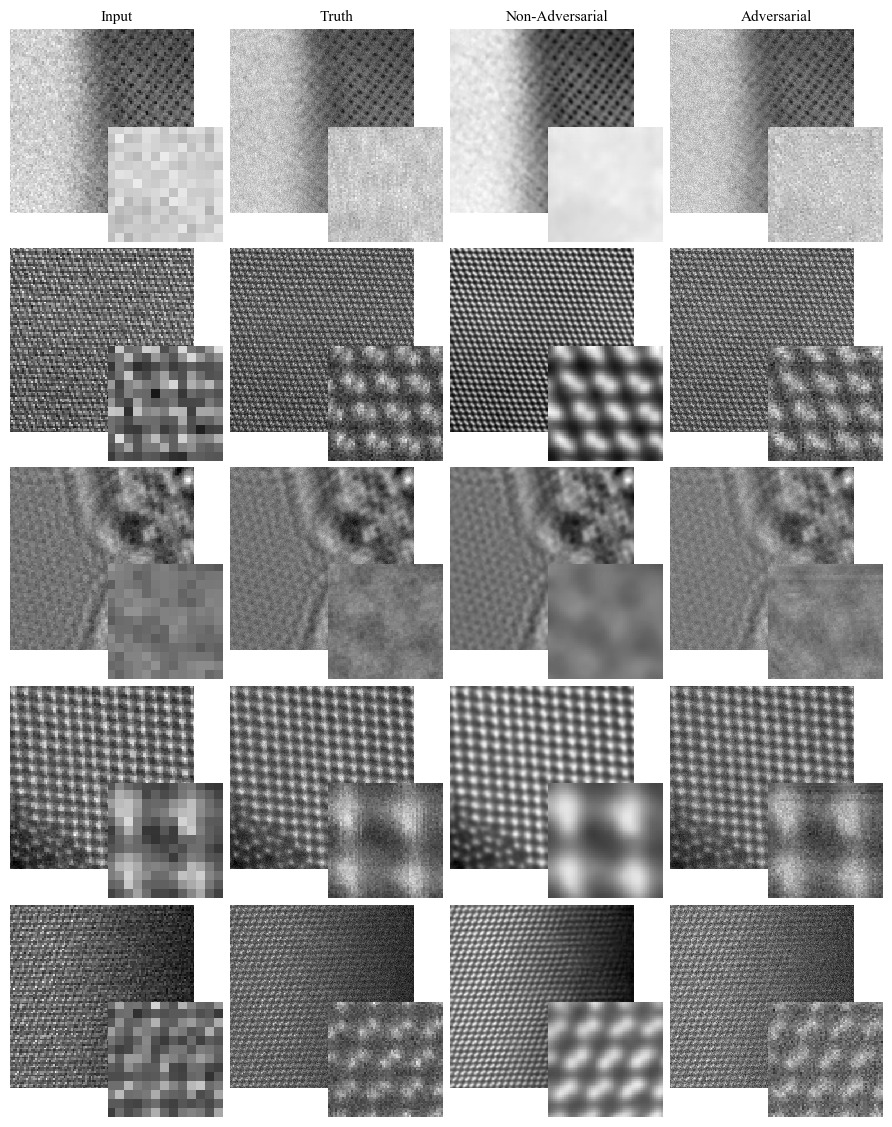}
\caption{ Adversarial and non-adversarial deep learning supersampling of test set 103$\times$103 STEM images to 512$\times$512. Adversarial completions have realistic noise characteristics and colouration whereas non-adversarial completions are blurry. Inputs are nearest neighbour upsampled to 512$\times$512 and enlarged 64$\times$64 regions from the top left of each image are inset to ease comparison. }
\label{examples-1}
\end{figure*}

\begin{figure*}[tbp!]
\centering
\includegraphics[width=0.97\textwidth]{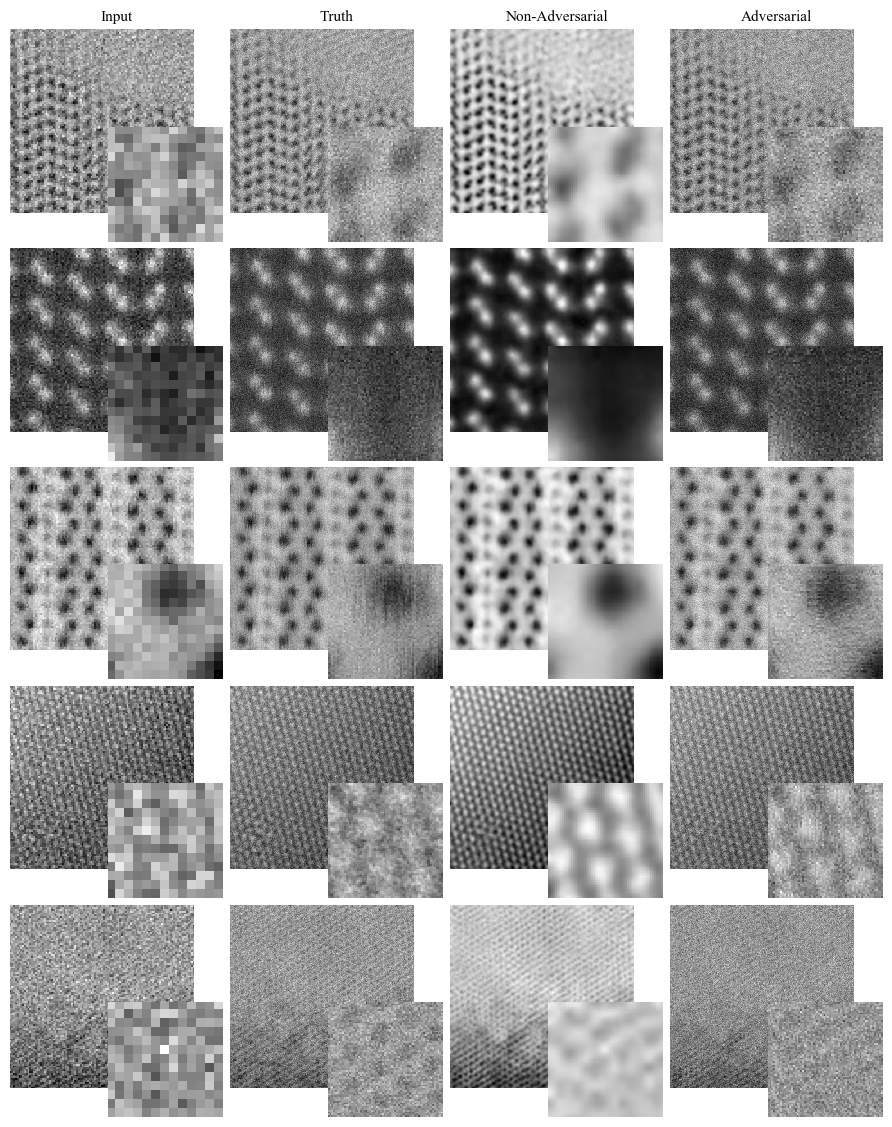}
\caption{ More adversarial and non-adversarial deep learning supersampling of test set 103$\times$103 STEM images to 512$\times$512. Adversarial completions have realistic noise characteristics and colouration whereas non-adversarial completions are blurry. Inputs are nearest neighbour upsampled to 512$\times$512 and enlarged 64$\times$64 regions from the top left of each image are inset to ease comparison. }
\label{examples-2}
\end{figure*}

\begin{figure*}[tbp!]
\centering
\includegraphics[width=0.97\textwidth]{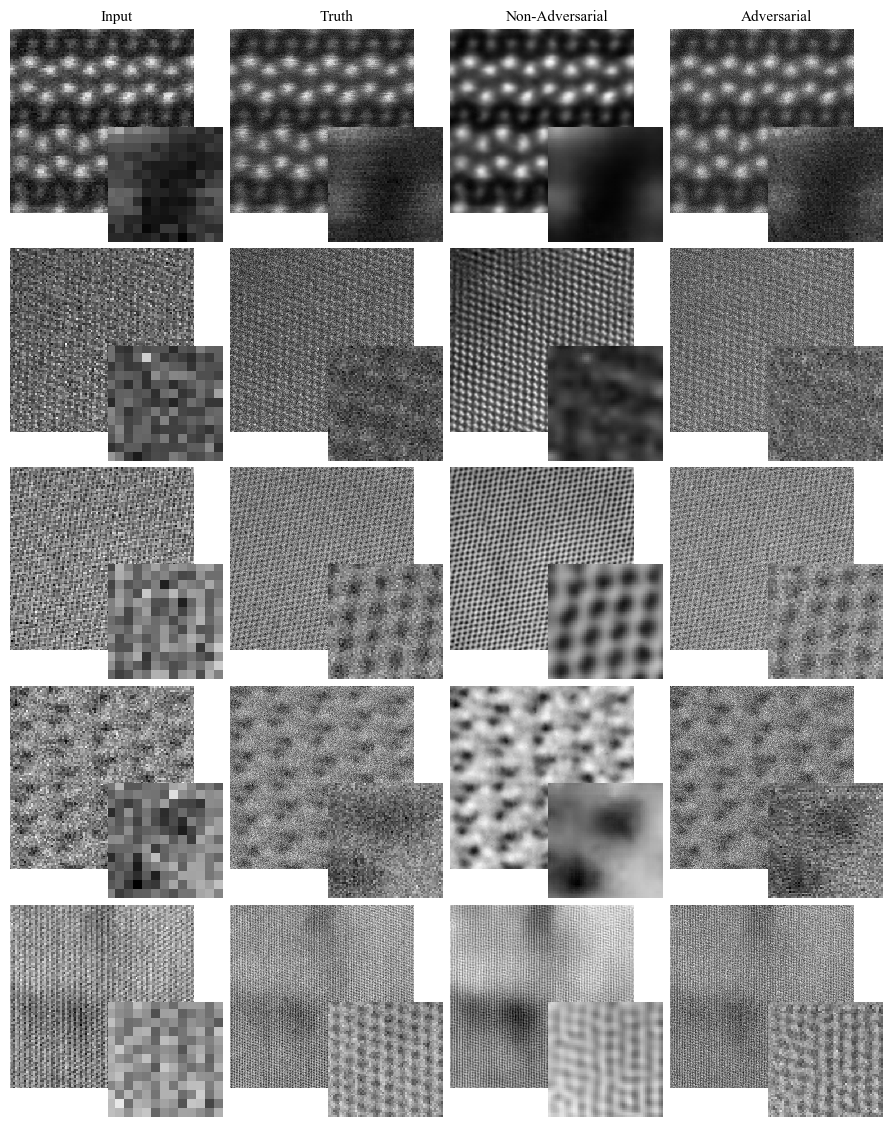}
\caption{ More adversarial and non-adversarial deep learning supersampling of test set 103$\times$103 STEM images to 512$\times$512. Adversarial completions have realistic noise characteristics and colouration whereas non-adversarial completions are blurry. The bottom row shows a failure case where detail is too fine. Inputs are nearest neighbour upsampled to 512$\times$512 and enlarged 64$\times$64 regions from the top left of each image are inset to ease comparison. }
\label{examples-3}
\end{figure*}

\begin{figure*}[tbp!]
\centering
\includegraphics[width=0.97\textwidth]{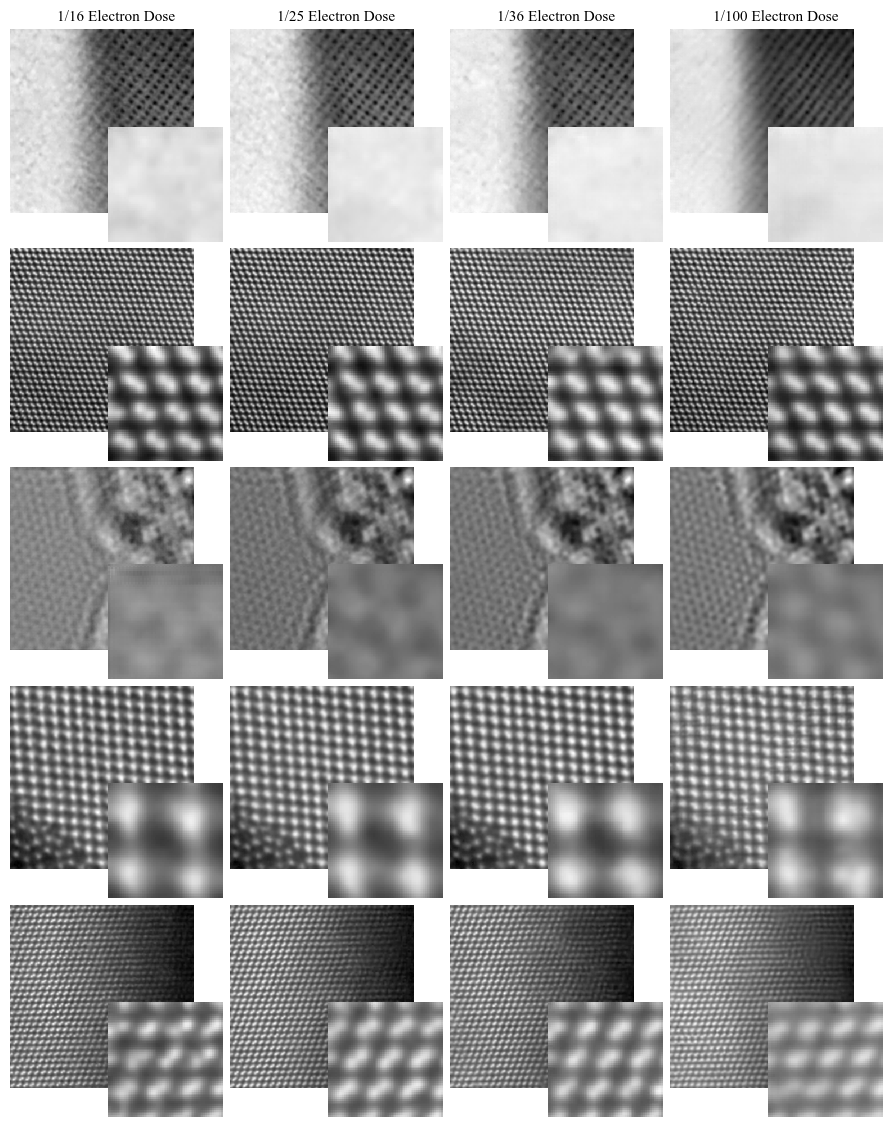}
\caption{ Non-adversarial deep learning supersampling of test set STEM images with 1/16, 1/25, 1/36 and 1/100 coverage. Supersampling becomes less accurate as coverage decreases. Enlarged 64$\times$64 regions from the top left of each image are inset to ease comparison. }
\label{different_doses-1}
\end{figure*}

\begin{figure*}[tbp!]
\centering
\includegraphics[width=0.97\textwidth]{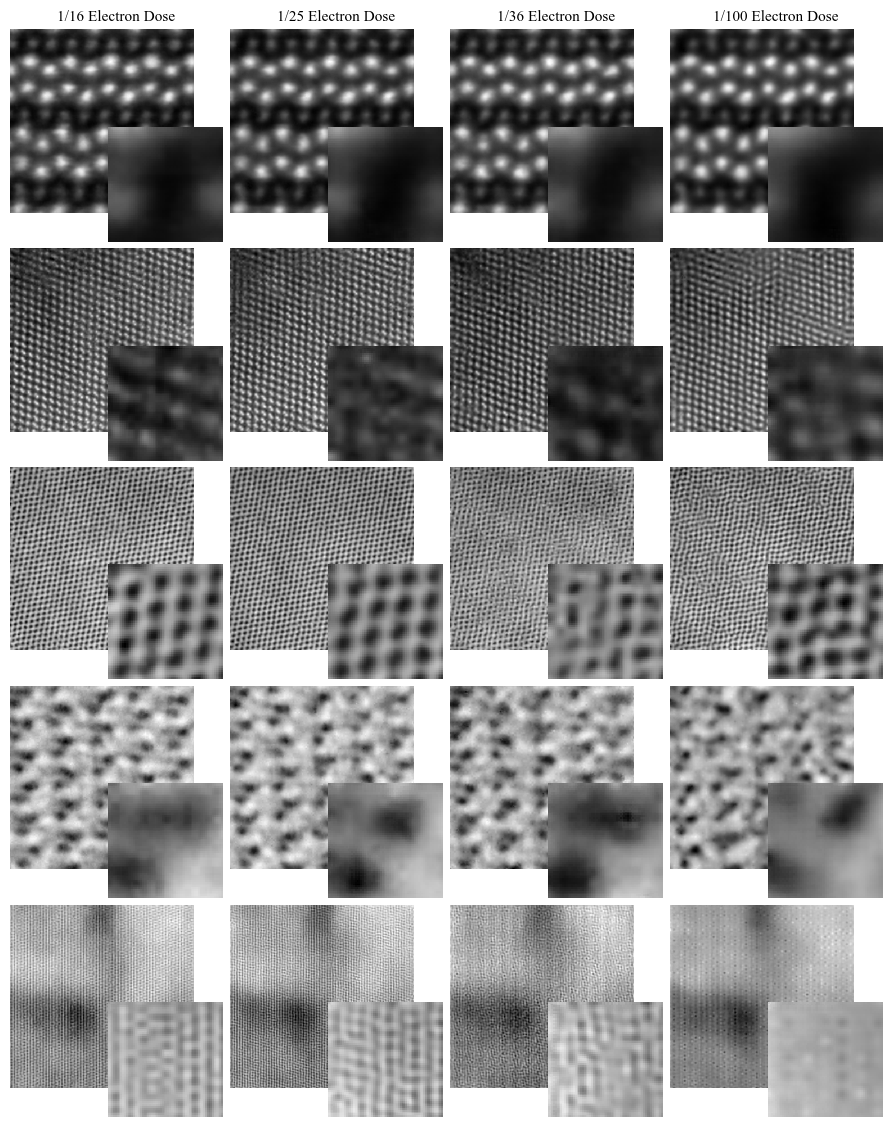}
\caption{ Non-adversarial deep learning supersampling of test set STEM images with 1/16, 1/25, 1/36 and 1/100 coverage. Supersampling becomes less accurate as coverage decreases. Enlarged 64$\times$64 regions from the top left of each image are inset to ease comparison. }
\label{different_doses-2}
\end{figure*}

\begin{figure*}[tbp!]
\centering
\includegraphics[width=0.97\textwidth]{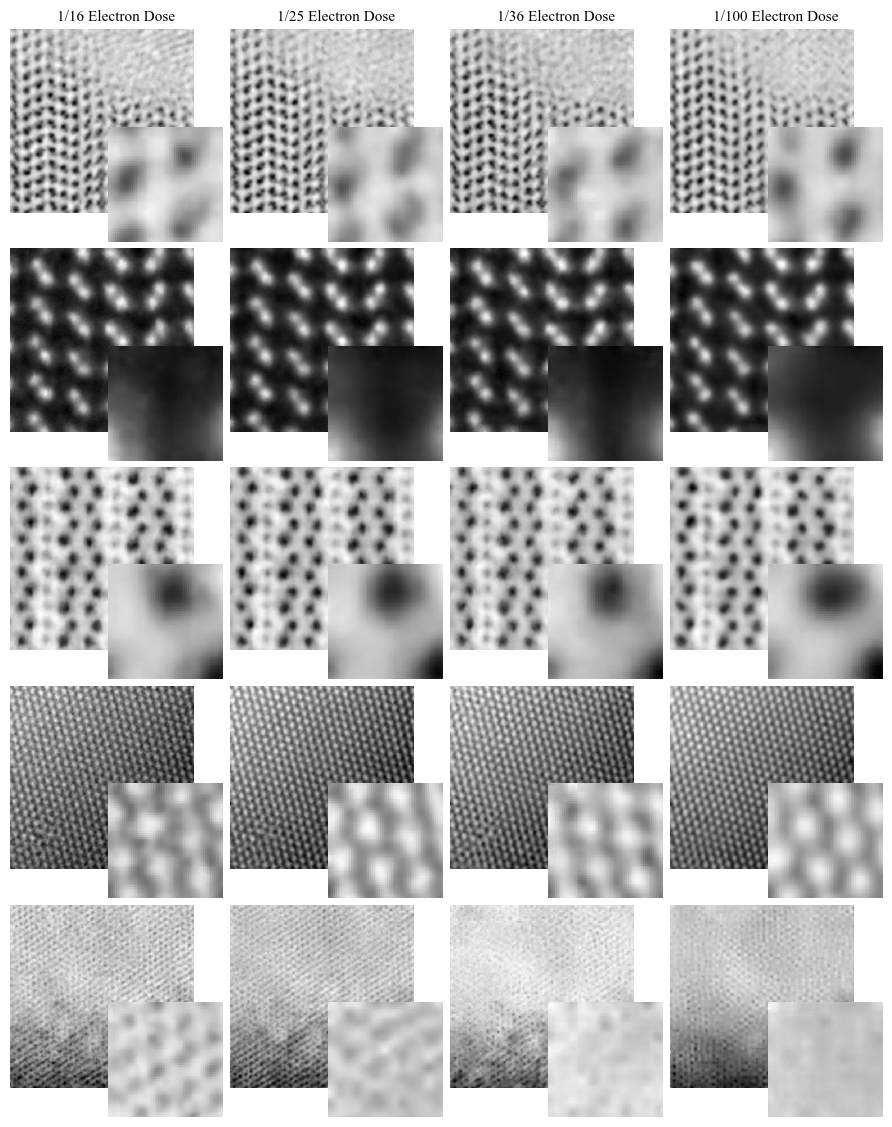}
\caption{ Non-adversarial deep learning supersampling of test set STEM images with 1/16, 1/25, 1/36 and 1/100 coverage. Supersampling becomes less accurate as coverage decreases. Enlarged 64$\times$64 regions from the top left of each image are inset to ease comparison. }
\label{different_doses-3}
\end{figure*}

\begin{figure*}[tbp!]
\centering
\includegraphics[width=0.97\textwidth]{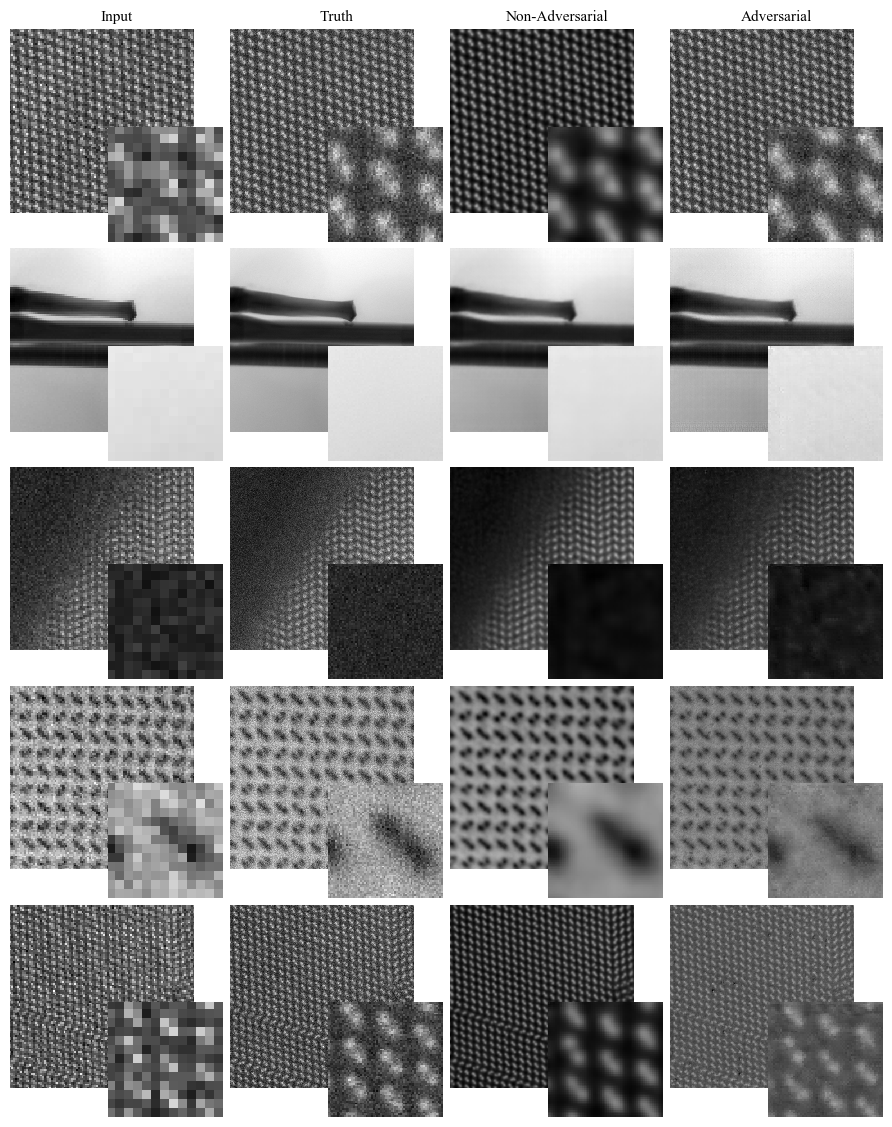}
\caption{ Adversarial and non-adversarial deep learning supersampling of test set 103$\times$103 STEM images to 512$\times$512. Adversarial completions have realistic noise characteristics and colouration whereas non-adversarial completions are blurry. Inputs are nearest neighbour upsampled to 512$\times$512 and enlarged 64$\times$64 regions from the top left of each image are inset to ease comparison. }
\label{unified_examples-1}
\end{figure*}

\begin{figure*}[tbp!]
\centering
\includegraphics[width=0.97\textwidth]{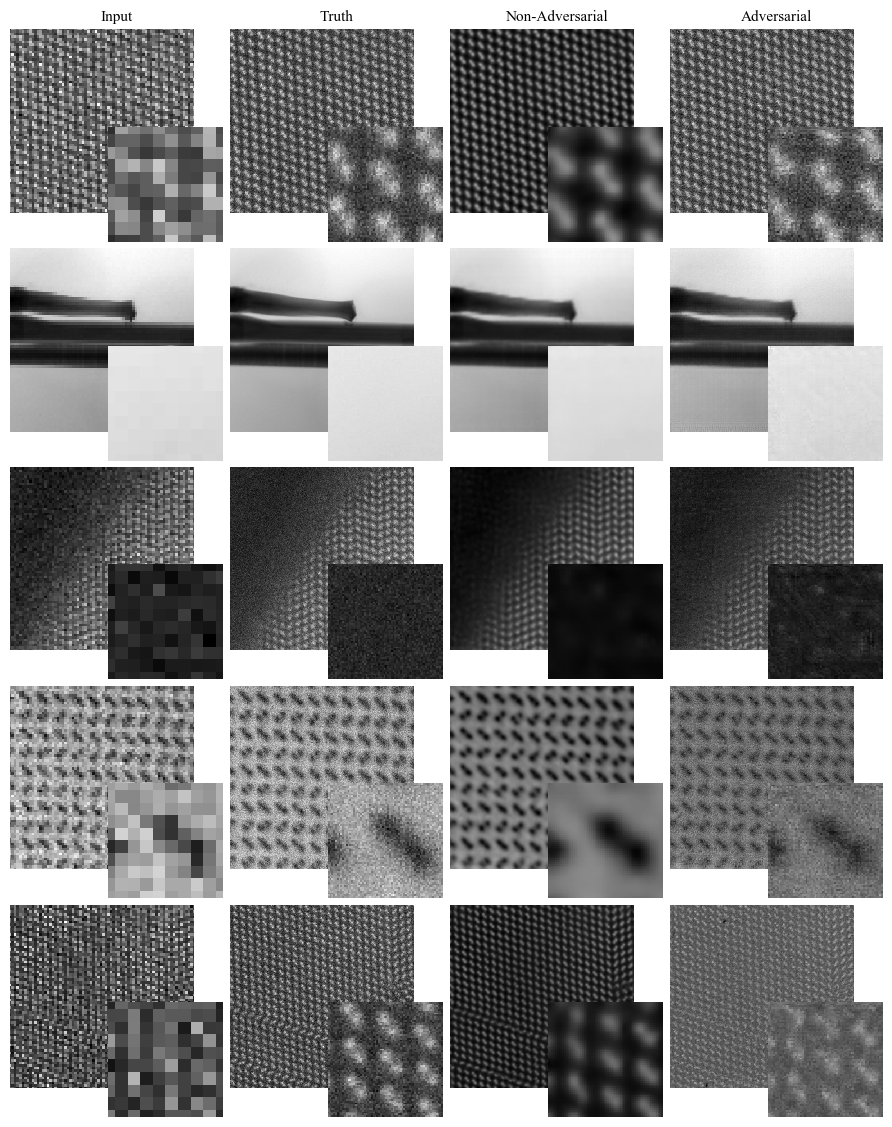}
\caption{ Adversarial and non-adversarial deep learning supersampling of test set 74$\times$74 STEM images to 512$\times$512. Adversarial completions have realistic noise characteristics and colouration whereas non-adversarial completions are blurry. Inputs are nearest neighbour upsampled to 512$\times$512 and enlarged 64$\times$64 regions from the top left of each image are inset to ease comparison. }
\label{unified_examples-2}
\end{figure*}

\begin{figure*}[tbp!]
\centering
\includegraphics[width=0.97\textwidth]{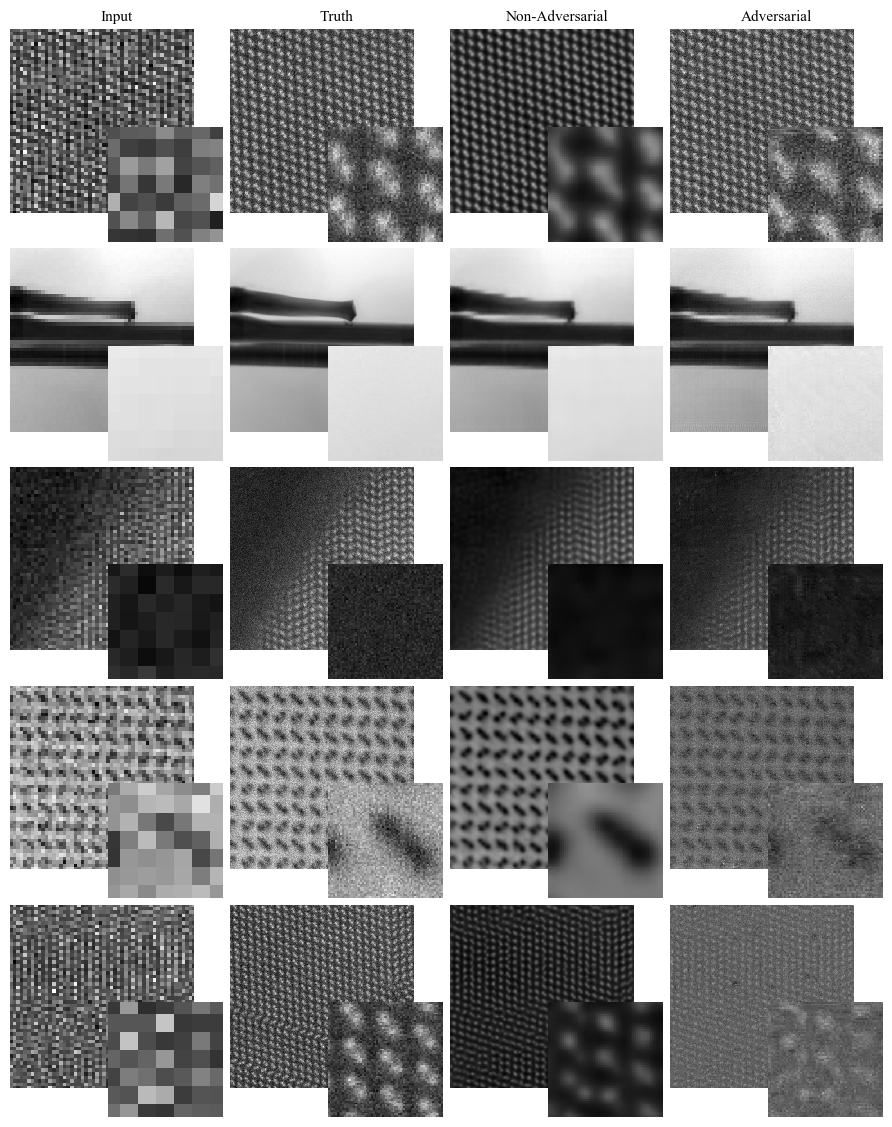}
\caption{ Adversarial and non-adversarial deep learning supersampling of test set 52$\times$52 STEM images to 512$\times$512. Adversarial completions have realistic noise characteristics and colouration whereas non-adversarial completions are blurry. The bottom row shows a failure case where detail is too fine. Inputs are nearest neighbour upsampled to 512$\times$512 and enlarged 64$\times$64 regions from the top left of each image are inset to ease comparison. }
\label{unified_examples-3}
\end{figure*}

Sheets of examples are appended. Comparisons of individual networks for 1/16, 1/25, 1/36, and 1/100 px coverage are in figs.~\ref{different_doses-1}-\ref{different_doses-3}. Their insets show that networks have higher errors at low coverage, especially for crops with fine detail. Example comparisons for non-adversarial and adversarial training for a 1/25 coverage generator are in figs.~\ref{examples-1}-\ref{examples-3}. They show that adversarial training produces images with realistic noise characteristics and colouration. Similar comparisons are made for a unified generator at 1/25, 1/49, and 1/100 px coverage in fig.~\ref{unified_examples-1}, fig.~\ref{unified_examples-2}, and fig.~\ref{unified_examples-3}, respectively. 

A 5$\times$5 symmetric Gaussian kernel with a 2.5 px standard deviation applied to target outputs is deconvoluted by gradient descent from 1/49 px non-adversarial outputs in fig.~\ref{deconv_panel}. Implicitly, non-adversarial networks learn to remove noise, similar to a denoising autoencoder\cite{vincent2008extracting}, so we aimed to deconvolute to less noisy outputs. However, while deconvolutions are less noisy than STEM images and remove blur, they also emphasize high-frequency grid-like spatial variation suppressed by Gaussian kernels. As shown in fig.~\ref{deconv_panel}, deconvolutional spatial error variation can be avoided by training without blurring outputs. Alternatively, high frequency spikes in Fourier transforms caused be gridlike artefacts could be divided by their average amplitudes or otherwise processed. 

\section{One-Stage Generator}

\noindent For generators trained for a single coverage, upsampling images to 512$\times$512 inefficiently utilizes initial convolutional layers as most input pixels are redundant. As a result, we heuristically implemented a one-stage generator, $G_\text{one-stage}$, shown in fig.~\ref{gen-1-step} for 1/25 coverage. This is intended to be a baseline to compare our two-stage generator against and a starting point for future single-coverage experiments. A comparison of one-stage and two-stage generator learning curves in fig.~\ref{learning_curves} for 1/25 px coverage shows that two-stage generator MSEs are lower. However, the two-stage generator received 5$\times$10$^5$ training iterations; compared to 10$^5$ for the one-stage generator, so the comparison may be uncharitable.

Before each convolutional layer, ReLU\cite{nair2010rectified} non-linearities are applied before batch normalization\cite{ioffe2015batch}, except before the first where no non-linearity is applied. Similar to single-coverage two-stage generator training, 512$\times$512 STEM crops were nearest neighbour downsampled to 103$\times$103 and the generator was non-adversarially optimized to minimize an ALRC MSE loss

\begin{equation}
L_\text{one-stage} = \text{ALRC}\left(\lambda_\text{one-stage} \text{MSE}(G_\text{one-stage}(I),  I_\text{blur})\right),
\end{equation}
where we chose $\lambda_\text{one-stage} = 200$. 

Training was ADAM\cite{kingma2014adam} optimized with batch size 16 for $10^5$ iterations. We used an initial learning rate of 0.001, which was stepwise linearly decayed to 0.0005, then 0.00025, at 12500 and 25000 iterations, respectively. After 50000 iterations, batch normalization was frozen and first moment of the momentum decay, $\beta_1$, was decreased from 0.9 to 0.5. 

\section{Web Serving}

\noindent At Warwick, we are developing an internal web server to host neural networks. This will allow anyone in the electron microscopy group to readily use networks without local hardware acceleration or software dependencies. To reduce latency, it helps to make data transfers as small as possible. In addition to data compression, we therefore experimented with reducing image bits from signed 32 bit to signed 16 bit and, with scaling, to unsigned 8 bit. Root mean squared errors on 20000 test set images for network inputs at different bit sizes are tabulated in table~\ref{types}. 

\begin{table}[tbhp]
\centering
\begin{tabularx}{\columnwidth}{cYYY}
\hline
 & 8 bit int & 16 bit float & 32 bit float \\
\hline
Mean & 0.0381 & 0.0345 & 0.0343 \\
Std Dev & 0.1075 & 0.0457 & 0.0447 \\
\hline
\end{tabularx}
\captionof{table}{ Root means and standard deviations of unclipped mean squared errors for 20000 test set images typecast to 8-bit integers, 16-bit floats, and 32-bit floats before input to a 1/25 dose network. }
\label{types}
\end{table}

Decreasing image bits increases root means and standard deviations of MSEs. Root MSEs are less than 1\% higher for 16 bit; rather than 32 bit, so it may be pragmatic typecast to 16 bit to reduce latency. Note that we do not currently have plans to make our server publicly accessible. However, it is a possibility in the future.

\section{Applications}

\noindent Test set results show that DLSS can decrease scan coverage and total electron electron dose by over an order of magnitude with a 3-4\% root MSE, enabling new beam-sensitive applications. In practice, this error is small compared to typical STEM noise so DLSS performance is comparable to standard STEM imaging. DLSS can also decrease peak electron dose where beam spread overlaps by enabling increased probing location separation. 

Pixels in point-scan system images are subsets of pixels in higher resolution images. As a result, DLSS can be applied as a post-processing step to ordinary STEM images, potentially long after a sample has be lost, modified or destroyed. Nevertheless, DLSS image characteristics are based on input image characteristics so imaging conditions can be optimized for DLSS based on a live DLSS display. Alternatively, a scan system could be modified to skip scan points without modifying imaging conditions.

Optimizing imaging conditions for DLSS may reducing the quality of the input image. To compare spatial correlation in input and target images, we measured mean absolute Laplacians (MAL) for 103$\times$103 segments of 20000 512$\times$512 validation set crops, $\textit{MAL}_\text{seg}$, and after nearest neighbour downsampling crops to 103$\times$103, $\textit{MAL}_\text{near}$. The mean ratio, $\textit{MAL}_\text{seg}/\textit{MAL}_\text{near}$, of 0.93 indicates higher spatial correlation in target images; as expected if experimenters optimize imaging conditions to increase spatial correlation. 

For low-lose imaging, it may be necessary to apply DLSS at all times, including during alignment or modification of imaging conditions, to prevent a sample from being damaged. For DLSS, this is possible because output images are based on input image characteristics. In a contrast, a network trained to generalize from artificial noise\cite{fang2019deep} may learn to produce average images, which may be less correlated with input images. Training based on pairs of high and low resolution images\cite{de2019resolution} overcomes this limitation; however, collecting training data is more expensive, especially for multiple coverages. 

Unified DLSS input size is fixed for each coverage. However, a network can be tiled across an image; potentially processing all the tiles in a single batch for computational efficiency. To reduce higher errors at the edge of DLSS outputs, tiles can overlap so that edges may be discarded. Alternatively, networks can be trained for other output sizes.

Following \cite{fang2019deep}, we propose that DLSS can be used to go beyond the resolution limit of an instrument. However, DLSS image characteristics are based on input image characteristics and may appear blurred if the scan system point spread function\cite{lupini2011three} (PSF) is larger than the pixel size. In contrast, a network trained to generalize from artificial noise, including blurs, is less susceptible to producing blurred outputs. Blurring can be corrected if the PSF is known or can be estimated\cite{roels2016bayesian}. However, PSF measurement is often unrealistic as it varies as imaging conditions are varied. As a result, we are investigating PSF deconvolution with a GAN. 

By enabling probing location coverage to be decreased by over an order of magnitude, DLSS enables scan speeds to be increased proportionately. This may enable increased temporal resolution of dynamic materials, including polar nanoregions in relaxor ferroelectrics\cite{kumar2019situ, xie2012static}, atom motion\cite{aydin2011tracking}, nanoparticle nucleation\cite{hussein2018tracking}, and material interface dynamics\cite{chen2018atomic}. Faster scans could also reduce delay for experimenters, decreasing microscope time.

\section{Conclusion}

\noindent This paper shows that DLSS-STEM can decrease electron dose and scan time, and increase resolution by up to 100$\times$, with minimal information loss. A novel learning policy is proposed to train a unified two-stage GAN for multiple coverages and training data with varying signal-to-noise ratios and shown to achieve similar MSEs to networks trained for individual coverages. Detailed MSE characteristics are provided for unified and individual networks, including MSEs for each output pixel. In addition, a baseline one-stage generator has been developed as a starting point for future work on individual coverages and we investigate numerical precision for web serving. We expect our results to be generalizable to other point-scan systems.

\section{Source Code}

\noindent A TensorFlow\cite{abadi2016tensorflow} implementation of DLSS-STEM is available at \url{\GitHubLoc}. 


\bibliographystyle{ieeetr}
\bibliography{bibliography}

\begin{thebibliography}{10}

\bibitem{wolf2018stem}
S.~G. Wolf, E.~Shimoni, M.~Elbaum, and L.~Houben, ``Stem tomography in
  biology,'' in {\em Cellular Imaging}, pp.~33--60, Springer, 2018.

\bibitem{garcia2014analysis}
A.~Garcia, A.~M. Raya, M.~M. Mariscal, R.~Esparza, M.~Herrera, S.~I. Molina,
  G.~Scavello, P.~L. Galindo, M.~Jose-Yacaman, and A.~Ponce, ``Analysis of
  electron beam damage of exfoliated mos2 sheets and quantitative haadf-stem
  imaging,'' {\em Ultramicroscopy}, vol.~146, pp.~33--38, 2014.

\bibitem{sem_review}
A.~Mohammed and A.~Abdullah, ``Scanning electron microscopy (sem): A review,''
  01 2019.

\bibitem{tong2018scanning}
Y.-X. Tong, Q.-H. Zhang, and L.~Gu, ``Scanning transmission electron
  microscopy: A review of high angle annular dark field and annular bright
  field imaging and applications in lithium-ion batteries,'' {\em Chinese
  Physics B}, vol.~27, no.~6, p.~066107, 2018.

\bibitem{goodfellow2014generative}
I.~Goodfellow, J.~Pouget-Abadie, M.~Mirza, B.~Xu, D.~Warde-Farley, S.~Ozair,
  A.~Courville, and Y.~Bengio, ``Generative adversarial nets,'' in {\em
  Advances in neural information processing systems}, pp.~2672--2680, 2014.

\bibitem{wang2017high}
T.-C. Wang, M.-Y. Liu, J.-Y. Zhu, A.~Tao, J.~Kautz, and B.~Catanzaro,
  ``High-resolution image synthesis and semantic manipulation with conditional
  gans,'' {\em arXiv preprint arXiv:1711.11585}, 2017.

\bibitem{binev2012compressed}
P.~Binev, W.~Dahmen, R.~DeVore, P.~Lamby, D.~Savu, and R.~Sharpley,
  ``Compressed sensing and electron microscopy,'' in {\em Modeling Nanoscale
  Imaging in Electron Microscopy}, pp.~73--126, Springer, 2012.

\bibitem{jmede2019partialscan}
J.~M. Ede and R.~Beanland, ``Partial scan electron microscopy with deep
  learning,'' {\em arXiv preprint arXiv:1905.13667}, 2019.

\bibitem{li2018compressed}
X.~Li, O.~Dyck, S.~V. Kalinin, and S.~Jesse, ``Compressed sensing of scanning
  transmission electron microscopy (stem) with nonrectangular scans,'' {\em
  Microscopy and Microanalysis}, vol.~24, no.~6, pp.~623--633, 2018.

\bibitem{reed2019electrostatic}
B.~Reed, A.~Moghadam, R.~Bloom, S.~Park, A.~Monterrosa, P.~Price, C.~Barr,
  S.~Briggs, K.~Hattar, J.~McKeown, {\em et~al.}, ``Electrostatic subframing
  and compressive-sensing video in transmission electron microscopy,'' {\em
  Structural Dynamics}, vol.~6, no.~5, p.~054303, 2019.

\bibitem{anderson2013sparse}
H.~S. Anderson, J.~Ilic-Helms, B.~Rohrer, J.~Wheeler, and K.~Larson, ``Sparse
  imaging for fast electron microscopy,'' in {\em Computational Imaging XI},
  vol.~8657, p.~86570C, International Society for Optics and Photonics, 2013.

\bibitem{sanders2018inpainting}
T.~Sanders and C.~Dwyer, ``Inpainting versus denoising for dose reduction in
  stem,'' {\em Microscopy and Microanalysis}, vol.~24, no.~S1, pp.~482--483,
  2018.

\bibitem{ferroni2016biological}
M.~Ferroni, A.~Signoroni, A.~Sanzogni, L.~Masini, A.~Migliori, L.~Ortolani,
  A.~Pezza, and V.~Morandi, ``Biological application of compressed sensing
  tomography in the scanning electron microscope,'' {\em Scientific reports},
  vol.~6, p.~33354, 2016.

\bibitem{guay2016compressed}
M.~D. Guay, W.~Czaja, M.~A. Aronova, and R.~D. Leapman, ``Compressed sensing
  electron tomography for determining biological structure,'' {\em Scientific
  reports}, vol.~6, p.~27614, 2016.

\bibitem{wu2018deep}
X.~Wu, R.-L. Li, F.-L. Zhang, J.-C. Liu, J.~Wang, A.~Shamir, and S.-M. Hu,
  ``Deep portrait image completion and extrapolation,'' {\em arXiv preprint
  arXiv:1808.07757}, 2018.

\bibitem{liu2018image}
G.~Liu, F.~A. Reda, K.~J. Shih, T.-C. Wang, A.~Tao, and B.~Catanzaro, ``Image
  inpainting for irregular holes using partial convolutions,'' in {\em
  Proceedings of the European Conference on Computer Vision (ECCV)},
  pp.~85--100, 2018.

\bibitem{yang2018deep}
W.~Yang, X.~Zhang, Y.~Tian, W.~Wang, and J.-H. Xue, ``Deep learning for single
  image super-resolution: A brief review,'' {\em arXiv preprint
  arXiv:1808.03344}, 2018.

\bibitem{de2019resolution}
K.~de~Haan, Z.~S. Ballard, Y.~Rivenson, Y.~Wu, and A.~Ozcan, ``Resolution
  enhancement in scanning electron microscopy using deep learning,'' {\em arXiv
  preprint arXiv:1901.11094}, 2019.

\bibitem{fang2019deep}
L.~Fang, F.~Monroe, S.~W. Novak, L.~Kirk, C.~Schiavon, B.~Y. Seungyoon,
  T.~Zhang, M.~Wu, K.~Kastner, Y.~Kubota, {\em et~al.}, ``Deep learning-based
  point-scanning super-resolution imaging,'' {\em bioRxiv}, p.~740548, 2019.

\bibitem{warwickem!}
J.~M. Ede, ``Stem crops dataset,'' {\em online:
  \url{https://warwick.ac.uk/fac/sci/physics/research/condensedmatt/microscopy/research/machinelearning}},
  2019.

\bibitem{ede2019adaptive}
J.~M. Ede and R.~Beanland, ``Adaptive learning rate clipping stabilizes
  learning,'' {\em arXiv preprint arXiv:1906.09060}, 2019.

\bibitem{thakallapelli2019alternating}
A.~Thakallapelli and S.~Kamalasadan, ``Alternating direction method of
  multipliers (admms) based distributed approach for wide-area control,'' {\em
  IEEE Transactions on Industry Applications}, vol.~55, no.~3, pp.~3215--3227,
  2019.

\bibitem{chen2006new}
J.~Chen, J.~Benesty, Y.~Huang, and S.~Doclo, ``New insights into the noise
  reduction wiener filter,'' {\em IEEE Transactions on audio, speech, and
  language processing}, vol.~14, no.~4, pp.~1218--1234, 2006.

\bibitem{kingma2014adam}
D.~P. Kingma and J.~Ba, ``Adam: A method for stochastic optimization,'' {\em
  arXiv preprint arXiv:1412.6980}, 2014.

\bibitem{salimans2016weight}
T.~Salimans and D.~P. Kingma, ``Weight normalization: A simple
  reparameterization to accelerate training of deep neural networks,'' in {\em
  Advances in Neural Information Processing Systems}, pp.~901--909, 2016.

\bibitem{hoffer2018norm}
E.~Hoffer, R.~Banner, I.~Golan, and D.~Soudry, ``Norm matters: efficient and
  accurate normalization schemes in deep networks,'' in {\em Advances in Neural
  Information Processing Systems}, pp.~2160--2170, 2018.

\bibitem{miyato2018spectral}
T.~Miyato, T.~Kataoka, M.~Koyama, and Y.~Yoshida, ``Spectral normalization for
  generative adversarial networks,'' {\em arXiv preprint arXiv:1802.05957},
  2018.

\bibitem{nair2010rectified}
V.~Nair and G.~E. Hinton, ``Rectified linear units improve restricted boltzmann
  machines,'' in {\em Proceedings of the 27th international conference on
  machine learning (ICML-10)}, pp.~807--814, 2010.

\bibitem{maas2013rectifier}
A.~L. Maas, A.~Y. Hannun, and A.~Y. Ng, ``Rectifier nonlinearities improve
  neural network acoustic models,'' in {\em Proc. icml}, vol.~30, p.~3, 2013.

\bibitem{vincent2008extracting}
P.~Vincent, H.~Larochelle, Y.~Bengio, and P.-A. Manzagol, ``Extracting and
  composing robust features with denoising autoencoders,'' in {\em Proceedings
  of the 25th international conference on Machine learning}, pp.~1096--1103,
  ACM, 2008.

\bibitem{ioffe2015batch}
S.~Ioffe and C.~Szegedy, ``Batch normalization: Accelerating deep network
  training by reducing internal covariate shift,'' {\em arXiv preprint
  arXiv:1502.03167}, 2015.

\bibitem{lupini2011three}
A.~R. Lupini and N.~De~Jonge, ``The three-dimensional point spread function of
  aberration-corrected scanning transmission electron microscopy,'' {\em
  Microscopy and Microanalysis}, vol.~17, no.~5, pp.~817--826, 2011.

\bibitem{roels2016bayesian}
J.~Roels, J.~Aelterman, J.~De~Vylder, H.~Luong, Y.~Saeys, and W.~Philips,
  ``Bayesian deconvolution of scanning electron microscopy images using
  point-spread function estimation and non-local regularization,'' in {\em 2016
  38th Annual International Conference of the IEEE Engineering in Medicine and
  Biology Society (EMBC)}, pp.~443--447, Ieee, 2016.

\bibitem{kumar2019situ}
A.~Kumar, R.~Dhall, and J.~M. LeBeau, ``In situ ferroelectric domain dynamics
  probed with differential phase contrast imaging,'' {\em Microscopy and
  Microanalysis}, vol.~25, no.~S2, pp.~1838--1839, 2019.

\bibitem{xie2012static}
L.~Xie, Y.~Li, R.~Yu, Z.~Cheng, X.~Wei, X.~Yao, C.~Jia, K.~Urban, A.~Bokov,
  Z.-G. Ye, {\em et~al.}, ``Static and dynamic polar nanoregions in relaxor
  ferroelectric ba (ti 1- x sn x) o 3 system at high temperature,'' {\em
  Physical Review B}, vol.~85, no.~1, p.~014118, 2012.

\bibitem{aydin2011tracking}
C.~Aydin, J.~Lu, A.~J. Liang, C.-Y. Chen, N.~D. Browning, and B.~C. Gates,
  ``Tracking iridium atoms with electron microscopy: first steps of metal
  nanocluster formation in one-dimensional zeolite channels,'' {\em Nano
  letters}, vol.~11, no.~12, pp.~5537--5541, 2011.

\bibitem{hussein2018tracking}
H.~E. Hussein, R.~J. Maurer, H.~Amari, J.~J. Peters, L.~Meng, R.~Beanland,
  M.~E. Newton, and J.~V. Macpherson, ``Tracking metal electrodeposition
  dynamics from nucleation and growth of a single atom to a crystalline
  nanoparticle,'' {\em ACS nano}, vol.~12, no.~7, pp.~7388--7396, 2018.

\bibitem{chen2018atomic}
S.~Chen, L.~Wang, R.~Shao, J.~Zou, R.~Cai, J.~Lin, C.~Zhu, J.~Zhang, F.~Xu,
  J.~Cao, {\em et~al.}, ``Atomic structure and migration dynamics of
  mos2/lixmos2 interface,'' {\em Nano energy}, vol.~48, pp.~560--568, 2018.

\bibitem{abadi2016tensorflow}
M.~Abadi, P.~Barham, J.~Chen, Z.~Chen, A.~Davis, J.~Dean, M.~Devin,
  S.~Ghemawat, G.~Irving, M.~Isard, {\em et~al.}, ``Tensorflow: A system for
  large-scale machine learning.,'' in {\em OSDI}, vol.~16, pp.~265--283, 2016.

\end{thebibliography}

\clearpage

\begin{figure*}[h]
\vspace{1.5cm}
\centering
\includegraphics[width=0.97\textwidth]{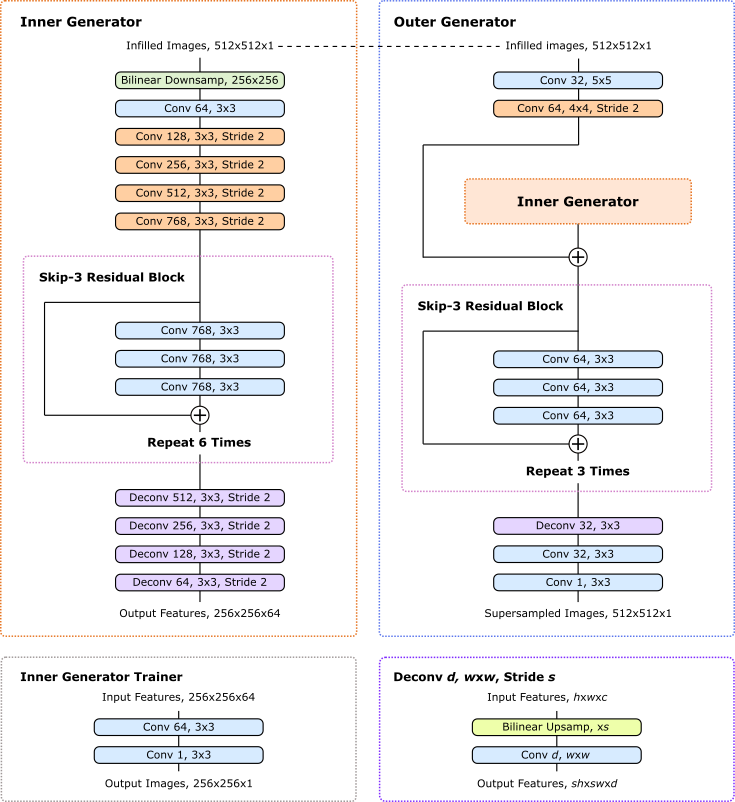}
\caption{ Two-stage generator that supersamples low resolution 103$\times$103$\times$1 images after nearest neighbour infilling to 512$\times$512$\times$1. A dashed line indicates that the same image is input to the inner and outer generator. Large scale features developed by the inner generator are locally enhanced by the outer generator and turned into images. An auxiliary inner generator trainer restores images from inner generator features to provide direct feedback.}
\label{gen-2-step}
\vspace{1.5cm}
\end{figure*}

\clearpage

\section{Detailed Architecture}\label{sec:architecture}

\begin{figure}[H]
\centering
\includegraphics[width=0.45\columnwidth]{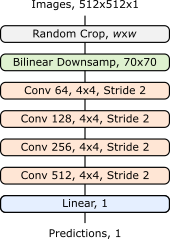}
\caption{ Discriminators examine random $w$$\times$$w$ crops to predict whether complete scans are real or generated. Adversarial generators are trained by multiple discriminators with different $w$. }
\label{discr}
\end{figure}

\noindent Network architecture from \cite{jmede2019partialscan} is summarized in this section. Generator and inner generator trainer architecture is shown in fig.~\ref{gen-2-step}. Discriminator architecture is shown in fig.~\ref{discr}. The components in our networks are

\vspace{\extraspace}
\noindent \textbf{Bilinear Downsamp, \textit{w}x\textit{w}:} This is an extension of linear interpolation in one dimension to two dimensions. It is used to downsample images to $w$$\times$$w$.

\vspace{\extraspace}
\noindent \textbf{Bilinear Upsamp, x\textit{s}:} This is an extension of linear interpolation in one dimension to two dimensions. It is used to upsample images by a factor of $s$.

\vspace{\extraspace}
\noindent \textbf{Conv \textit{d}, \textit{w}x\textit{w}, Stride, \textit{x}:} Convolution with a square kernel of width, $w$, that outputs $d$ feature channels. If the stride is specified, convolutions are only applied to every $x$th spatial element of their input, rather than to every element. Striding is not applied depthwise.

\vspace{\extraspace}
\noindent \textbf{Linear, \textit{d}:} Flatten input and fully connect it to $d$ feature channels.

\vspace{\extraspace}
\noindent \textbf{Random Crop, \textit{w}x\textit{w}:} Randomly sample a $w$$\times$$w$ spatial location using an external probability distribution. 

\vspace{\extraspace}
\noindent \textbf{\circled{+}:} Circled plus signs indicate residual connections where incoming tensors are added together. These help reduce signal attenuation and allow the network to learn perturbative transformations more easily.

All generator convolutions are followed by running mean-only batch normalization then ReLU activation, except output convolutions. All discriminator convolutions are followed by slope 0.2 leaky ReLU activation.

\section{Acknowledgements}

\noindent This research was funded by EPSRC grant EP/N035437/1.

\end{document}